\documentclass[sn-mathphys,Numbered]{sn-jnl}

\usepackage{microtype} 

\usepackage{graphicx}%
\usepackage{multirow}%
\usepackage{amsmath,amssymb,amsfonts}%
\usepackage{amsthm}%
\usepackage{mathrsfs}%
\usepackage[title]{appendix}%
\usepackage{xcolor}%
\usepackage{textcomp}%
\usepackage{manyfoot}%
\usepackage{booktabs}%
\usepackage{algorithm}%
\usepackage{makecell}

\usepackage{lmodern}

\usepackage{cleveref}

\usepackage{listings}%

\usepackage{mathtools}

\usepackage[accsupp]{axessibility}

\usepackage{amsthm} 
\usepackage{etoolbox}

\DeclareMathOperator*{\argmin}{arg\,min}
\usepackage{dsfont}
\usepackage{comment}

\usepackage[outdir=figures/]{epstopdf}

\DeclarePairedDelimiterX{\abs}[1]{\lvert}{\rvert}{%
  \ifblank{#1}{\,\cdot\,}{#1}
}   

\DeclarePairedDelimiterX\norm[1]\lVert\rVert{%
  \ifblank{#1}{\,\cdot\,}{#1}
}   

\DeclarePairedDelimiterX\Set[1]\{\}{%
  
  #1
}

\DeclarePairedDelimiterX\innerp[2]{\langle}{\rangle}{%
  \ifblank{#1}{\,\cdot\,}{#1} , \ifblank{#2}{\,\cdot\,}{#2}%
}

\DeclarePairedDelimiterX\braket[2]{\langle}{\rangle}%
  {#1\kern0.15ex\delimsize\vert\kern0.15ex\mathopen{}#2}

\DeclarePairedDelimiterX\ketbra[2]{\vert}{\vert}%
  {#1\kern0.15ex\delimsize\rangle\delimsize\langle\kern0.15ex\mathopen{}#2}

\DeclarePairedDelimiterX\sandwich[3]{\langle}{\rangle}%
  {#1\,\delimsize\vert\kern0.15ex\mathopen{}#2\kern0.15ex\delimsize\vert\kern0.15ex\mathopen{}#3}
\usepackage{algpseudocode}
\algrenewcommand \algorithmicrequire{\textbf{Input:}}

\algrenewcommand \algorithmicensure{\textbf{Goal:}}

\usepackage{physics}

\theoremstyle{thmstyleone}%

\usepackage{subfig}

\newcommand{\lin}{ \textrm{lin} }
\newcommand{\quadr}{ \textrm{quad} }
\newcommand{\ord}{ \textrm{ord} }

\begin{document}

\title[Article Title]{Compensating connectivity restrictions in quantum annealers via splitting and linearization techniques}

\author*[1]{\fnm{Marcel} \sur{Seelbach Benkner} 
\orcid{https://orcid.org/0000-0003-4225-9793}
\,}\email{marcel.seelbach@eleqtron.com}

\author[1,4]{\fnm{Zorah} \sur{Lähner} \orcid{https://orcid.org/0000-0003-0599-094X}
\,}

\author[2]{\fnm{Vladislav} \sur{Golyanik}  
\orcid{https://orcid.org/0000-0003-1630-2006}
\,}

\author[3]{\fnm{Martin} \sur{Kliesch}  \orcid{https://orcid.org/0000-0002-8009-0549}
\,}

\author[1]{\fnm{Michael} \sur{Moeller} \orcid{https://orcid.org/0000-0002-0492-6527} 
\,}

\affil*[1]{\orgdiv{Naturwissenschaftlich-Technische 
Fakultät}, \orgname{Universität Siegen}, \orgaddress{\street{Hölderlinstraße 3}, \city{Siegen}, \postcode{57076},  \country{Germany}}}

\affil[2]{\orgdiv{MPI Informatik}, \orgname{Saarland Informatics Campus}, \orgaddress{\street{Stuhlsatzenhausweg E1 4}, \city{Saarbrücken}, \postcode{66123},  \country{Germany}}}

\affil[3]{\orgdiv{Institute for Quantum Inspired and Quantum Optimization}, \orgname{Hamburg University of Technology}, \orgaddress{\street{Blohmstraße 15}, \city{Hamburg}, \postcode{21079}, \country{Germany}}}

\affil[4]{\orgdiv{Lamarr Institute for Machine Learning and Artificial Intelligence}, \orgname{Rheinische Friedrich-Wilhelms-Universität Bonn}, \orgaddress{\street{Regina-Pacis-Weg 3}, \city{Bonn}, \postcode{53113}, \country{Germany}}}

\abstract{Current quantum annealing experiments often suffer from restrictions in connectivity in the sense that only certain qubits can be coupled to each other. The most common strategy to overcome connectivity restrictions so far is by combining multiple physical qubits into a logical qubit with higher connectivity, 
which is achieved by adding terms to the Hamiltonian. 
Practically, this strategy is implemented by finding 
a so-called minor embedding, which is in itself an NP-hard problem.
In this work, we present an iterative algorithm that does not need additional qubits but instead efficiently uses the available connectivity for different parts of the problem graph in every step. We present a weak monotonicity proof and benchmark our algorithm against the default minor-embedding algorithm on the D-Wave quantum annealer and multiple simple local search variants. While most of the experiments to compare the different iterative methods are performed with simulated annealing solvers, we also confirm the practicality of our method with experiments on the D-Wave Advantage quantum annealer.

}

\maketitle

\section{Introduction}\label{sec1}

Quantum annealing is a recently emerging optimization method that promises to outperform, e.g., simulated annealing in rugged energy landscapes.
The term quantum annealing is often used to refer to the current experiments motivated by adiabatic quantum computing, which may not achieve perfect adiabaticity \cite{AdiabaticQuantumComputingandQuantumAnnealing}. 
Due to this imperfection, 
the systems are not necessarily in the ground state throughout the whole evolution, but also go to low-energy excited states. 
While adiabatic quantum computing is under some theoretical assumptions polynomially equivalent to gate-based quantum computing \cite{aharonov2008adiabatic},
we are in this work mostly interested in quantum annealing as a heuristic solver for quadratic unconstrained optimization (QUBO) problems. 
These problems have the form
\begin{equation}
  \hat{x}:=  \argmin_{x \in \{0,1\}^n }  x^T Q x,
\label{eq:QUBO}
\end{equation}
where $Q\in \mathbb{R}^{n\times n } $ is a matrix of couplings, which we assume to be symmetric without restriction of generality. For QUBOs, there is also an equivalent formulation to the Ising problem 
\begin{equation}
    \hat{s}:=    \argmin_{s \in \{-1,1\}^n }  s^T A s +s^T b ,
        \label{eq:Ising}
\end{equation}
with $A= \frac{1}{4}Q$ and $b= \frac{1}{2}  Q  \mathbf{1} $, where $\mathbf{1} \in \mathbb{R}^n$ is a vector with only ones as entries.
QUBO problems can be solved with traditional algorithms, like simulated annealing, but the energy landscape can become very complicated, and the problem of finding the optimum is NP-hard in general. It is believed that quantum annealing can often handle these landscapes more efficiently and robustly \cite{das2005quantum,PhysRevX.6.031015}, and there are even problems where a quadratic 
speedup compared to the best foreseeable classical algorithm is mathematically proven \cite{roland2002quantum}.

To solve the minimization \eqref{eq:QUBO} with quantum annealing in a straightforward way, $n$ logical qubits are necessary.
However, most current quantum annealing experiments are still quite restricted in the number of qubits and thus in the problem size. The largest system is currently manufactured by the company D-Wave\footnote{https://www.dwavesys.com/} and uses superconducting flux qubits \cite{mcgeoch2014adiabatic}. Although the D-Wave Advantage machine has over 5000 qubits, it can only solve QUBO problems up to the size of $\approx 150$ variables
if $Q$ is a dense matrix \cite{dwavesysSolversMinorEmbedding}.
This overhead is due to physical limitations in connecting qubits: a non-zero entry in $Q$ requires a connection between the two corresponding qubits, but it is currently not possible to manufacture all qubits such that they are all connected. 
To overcome this limitation, multiple physical qubits are combined into a logical qubit with higher connectivity, with the drawback that the number of logical qubits is considerably lower than the physical ones. The process of computing which physical qubits need to be combined to form one logical qubit is called finding a \textit{minor embedding}.
If a coupling matrix $Q$ has a zero diagonal, then we can consider it to be the adjacency matrix of a weighted graph, which we also refer to as the problem graph.  
When solving multiple problems with the quantum annealer, the minor embedding of the problem graph onto the hardware graph is usually recomputed before the annealing takes place. For the physical realization of the minor embedding, one has to add terms to the Hamiltonian that enforce similar behavior within the so-called \textit{chain} of physical qubits that are combined into a logical qubit. The \textit{chain strength} describes how strongly unequal measurement values (in the computational basis) for qubits inside the chain are penalized.
The default way to calculate the minor embedding is described by \citet{Cai2014}. 

In this work, we introduce a hybrid method that does not make use of minor embeddings but instead optimizes only over a subset of edges of the problem graph that are directly present in the hardware graph. In other words, we use the hardware graph as a mask to pick edges of the problem graph for the quadratic terms of the subproblems and linearize the rest. In order to take care of all parts of the problem graph, one regularly changes the correspondence between the problem and the hardware graph via a random permutation. To build upon the computations of the last iteration before the permutation of the hardware graph, one introduces a damping term. While we did not find an easy recipe to predict for which cases our iterative method should be preferred over minor embedding methods, it is obvious that our method can be applied to larger problems and has less difficulty with highly connected graphs. Methods that generate subproblems by keeping some variables fixed and optimizing over the rest are an important comparison for our iterative algorithm. Whether these methods perform better, irrespective of possible minor embedding problems, is very dependent on the size of the subproblems. We also want to stress that the method presented here, as well as many subproblem methods \cite{raymond2023hybrid}, has the drawback that there is no guarantee of getting global optimal solutions even if the underlying hardware-restricted QUBO solver were perfect. 
However, we will also sketch a rough idea of how it might be possible to extend these methods for global optimization by adding stochastic terms that make the resulting method non-monotone.

In the following section, we will refer to some publications where computing minor embeddings plays a crucial role or where one iteratively solves subproblems. 
Then we introduce the new method in \cref{sec:Methods} and present some theoretical results about its weak monotonicity in \cref{sec:Proof}. 
Finally, in \cref{sec:Experiments}, we present experiments to extensively benchmark the presented method against a large neighborhood local search of various sub-problem sizes. We also show and analyze problems that cannot be addressed using minor embeddings (i.e.~of comparably large sizes). Next, we present the results of proof-of-concept implementations of our method on the actual D-Wave hardware.

\section{Related work}%
\label{sec:Rel}
Many articles in the literature aim at improving the calculation of minor embeddings or consider special cases of problem graphs. On the other hand, 
numerous works exist in the mathematical optimization
and operation research communities on how to solve large optimization problems in an iterative, piece-wise, and often heuristic manner. We will discuss now
how some of these works are connected to the optimization with D-Wave quantum annealers.

\paragraph{Hardware embeddings} 
The problem of finding a minor embedding naturally arises
for hardware-restricted quantum annealing solvers. For the D-Wave machines, this led to the necessity to research fast and practical
algorithms to calculate minor embeddings. In 2014, Cai et al. published a first heuristic for this problem, while most previous
research dealt with exact algorithms with calculable time complexity \cite{Cai2014}. 
Subsequent work 
often makes it possible to compute embeddings with 
fewer physical qubits even faster \cite{goodrich2018optimizing,patton2019efficiently}. 
Designing such heuristics is crucial for the success of quantum annealing via minor embeddings, as the task of finding an embedding for the typical hardware graphs in D-Wave machines is NP-hard 
\cite{LOBE2024114369}. 
This statement depends on the fact that, usually, there are defective qubits present which have to be accounted for. 
On the other hand, the question of whether a certain graph is embeddable can be answered in polynomial time \cite{ROBERTSON199565} if the hardware graph is fixed. 
Another option is to precompute minor embeddings, e.g.\ for cliques \cite{boothby2016fast} or for cubic lattices \cite{king2020performance}. 
However, if the actual problem to solve in the end has fewer edges than, e.g.\ the fully connected problem, one typically has worse performance compared to specially tailored embeddings.

\paragraph{Decomposition, preprocessing or iterative methods}
Besides minor embeddings, divide-and-conquer \cite{guerreschi2021solving} or decomposition approaches, e.g.\ \cite{ponce2023graph}, are common ways to deal with large QUBOs.
More specifically, Ref.~\cite{guerreschi2021solving} views the couplings of the QUBO as a graph that is divided into subgraphs such that the weights of the edges within the subgraph are strong compared to the weights of possible edges that go out of the subgraph - an approach called \textit{community detection}. 
Then, the resulting inter- and extra-community interactions are treated separately.

If one encounters Ising problems that are initially too large for the quantum annealer, one can also try to reduce the number of qubits with preprocessing techniques like in Ref.~\cite{thai2022fasthare}. The idea is to identify sets of vertices that will have the same value assigned to them in the final solution. By treating these sets as single vertices, one can set up a smaller, reduced problem. In Ref.~\cite{thai2022fasthare}, the preprocessing technique was applied to Max-Cut problems from the MQLib dataset \cite{DunningEtAl2018}.

Due to their equivalence to QUBOs, Max-Cut problems are often considered for benchmarking in quantum optimization. For example, in Ref.~\cite{Dupont2024QuantumOF} there is a particularly exhaustive comparison of different methods, including a Quantum Approximate Optimization Algorithm (QAOA) \cite{farhi2014quantum} method on a gate-based superconducting quantum computer.
 Furthermore, it is also possible to compute several problems on the quantum annealer in parallel \cite{pelofske2022parallel}.
Another publication where multiple decomposition methods are investigated is
\cite{bass2021optimizing}.
One insight that was gathered there is that it is often preferable not to recompute the minor embedding for every small subproblem in iterative algorithms.

The work most related to ours is Ref.~\cite{raymond2023hybrid}.
Here, the authors use the quantum annealer
with a large-neighborhood local-search hybrid algorithm.
They mostly consider lattice-like problems in which the embeddings are trivial or can be precomputed easily. We, on the other hand, focus more on datasets with diverse connectivity patterns or even fully connected problems: Our proposed method works independently of the connectivity structure of the problem instance.

\paragraph{Classical optimization and constraints }

Our method falls in the same category as many proximal gradient descent methods that were developed for 
composite optimization problems that occur in compressive sensing or image denoising. 
In particular, one obstruction to using 
simple gradient descent algorithms for these applications is that the objective often has a part that is not differentiable.
An excellent monograph to learn about proximal algorithms can be found in Parikh et al.~\cite{parikh2014proximal}. From this source, we will introduce the reader to several facts about this topic.
First, the definition of a \emph{proximal operator} is the following: it maps between real-valued functions as $f\mapsto \operatorname{prox}_{f}$ with
\begin{equation}
    \operatorname{prox}_{f}(v)= \argmin_{x\in \mathbb{R}^n} \left( f(x)+ \frac{1}{2} || x-v ||_2^2\right). 
\end{equation}
This operator can be viewed as a generalization of a projection on a set. 
As already mentioned, one is interested in composite optimization problems of the form
\begin{equation}
    \min_{x\in \mathbb{R}^n} f(x)+ g(x),
\end{equation}
where $f$ and $g$ are closed, proper convex functions. $f:\mathbb{R}^n\to \mathbb{R} $ is assumed to be differentiable, while $g:\mathbb{R}^n\to \mathbb{R} \cup \{+\infty\}$ does not have to be differentiable and is often thought of as a way to incorporate constraints via a characteristic function of a set. 
The proximal gradient descent 
algorithm for this, formulated as in Ref.~\cite{parikh2014proximal}, is given by
\begin{equation}
    x^{(k+1)}= \text{prox}_{\lambda^{(k)} g}\left( 
x^{(k)} -\lambda^{(k)} \nabla f(x^{(k)})  \right)
\end{equation}
with an iteration dependent step size parameter $\lambda^{(k)}$. Inserting the formula for the proximal operator yields 
\begin{align}
     x^{(k+1)}&= \argmin_{x\in \mathbb{R}^n} \left( \lambda^{(k)} g(x) + \frac{1}{2} || x-(x^{(k)} -\lambda^{(k)} \nabla f(x^{(k)}) ) ||_2^2\right) \\
     &= \argmin_{x\in \mathbb{R}^n} \left(\lambda^{(k)} g(x) + \lambda^{(k)} \langle   \nabla f(x^{(k)})  ,x-x^{(k)} \rangle+ \frac{1}{2} || x-x^{(k)} ||_2^2 \right. \nonumber\\  & \left.+\frac{1}{2} || \lambda^{(k)} \nabla f(x^{(k)})  ||_2^2\right), \label{eq:ProxIt}
\end{align}
where the last summand is irrelevant for the $\argmin$. In the context of compressive sensing, the function $g(x)$ is often $||x||_1$ to encourage sparsity of a solution vector. 
Since the proximal operator of the 1-norm reduces the absolute value of the individual components, popular algorithms that tackle this problem with proximal gradient descent are called the Iterative Shrinkage Threshold Algorithm (ISTA) 
or Fast Iterative Shrinkage Threshold Algorithm
(FISTA) (s.\ \cite{beck2009fast} and the references therein). 
An extension to FISTA for the non-convex case is studied in Ref.~\cite{ochs2019adaptive}. We are not aware of extensions of proximal gradient descent to discrete optimization that are relevant here.

The Alternating Direction Method of Multipliers (ADMM) is also interesting in our context. Here, one also looks at composite problems, but $f$ and $g$ can be non-differentiable and can incorporate constraints. This method is useful if $f+g$ is hard to optimize directly, but the individual parts are easy to optimize.

The publication \cite{gambella2020multiblock} investigates how to use QUBO solvers for mixed binary optimization problems and applies ADMM for this purpose.  
In Ref.~\cite{wu2018ell}, ADMM is also used for integer programming and binary quadratic programming as a special case. One main idea here is that the binary vector constraints are written as the intersection of a hypersphere and a hypercube.

Although this is related to our use of proximal gradient descent algorithm, they aim to extend 
QUBO-solvers to mixed binary optimization solvers and are not concerned with hardware restrictions in quantum annealers.
Another method concerned with 
iteratively solving constrained optimization problems on the D-Wave quantum annealer is developed in Ref.~\cite{yurtsever2022q}.
If one considers a concrete class of problems to be solved with the D-Wave machines, one is well advised to spend some time on considering efficient ways to formulate the problem as a QUBO problem \cite{fowler2017improved,benkner2020adiabatic,meli2025qucoop}. 
In the easiest case, this can be done through
eliminating variables by solving for them in one of the equality constraints and inserting the solution.

Furthermore, there are even more relevant classical methods to solve QUBO problems.
A relaxation-based method for solving binary quadratic unconstrained problems with some similarities to our method can be seen in Ref.~\cite{olsson2008improved}.
In Merz and Freisleben \cite{merz2002greedy}, a greedy algorithm as well as 1-Opt and the generalization k-Opt are evaluated on QUBO instances. The k-Opt algorithm goes in each iteration to the binary vector that differs from the current binary vector by no more than $k$ entries and has the best energy among these.
We will compare against k-Opt algorithms in the experiments section.

\section{Methods}\label{sec:Methods}

In this section, we describe the proposed iterative optimization method.
Our method does not need a minor embedding and is motivated by a weak monotonicity proof 
that guarantees convergence (s.  \cref{sec:weak_mono}).
Additionally, we investigate how multiple known methods are related to ours.
In particular, projected gradient descent (s. \cref{sec:RelGradDesc}) and local search in the sense of 
optimizing over subproblems while fixing the other variables (s. \cref{sec:FixVar}) are relevant. In \cref{sec:SAlikeReg}, we present a regularization for optimizing over subproblems with a concept similar to combining simulated annealing and local search as in Ref.~\cite{martin1996combining}.

\subsection{Linearization and damping term (Splitting)}\label{sec:LiDa}
The main idea of our method is to split the coupling matrix $A$ from the Ising problem \eqref{eq:Ising} into two parts, $A= A_{\quadr} + A_{\lin}$, keep the original (quadratic) costs for $A_{\quadr}$, and linearize the costs involving $A_{\lin}$.

Therefore, we want to optimize over quadratic functions of the form 
\begin{equation}
    E( s) = s^TA s= s^T A_{\quadr} s +  s^T A_{\lin} s, \label{eq:quadEn}
\end{equation}
where $A$, $A_{\quadr}$, and $A_{\lin}$ are symmetric matrices in $\mathbb{R}^{n\times n}$. The matrix $A_{\quadr}$ could be equal to $A$ in all the entries that correspond to a hardware graph from some quantum annealer architecture. A bias term $b^Ts$ can, of course, also be present, but we omit it here for better readability. 
We propose to use the proximal gradient descent algorithm \eqref{eq:ProxIt} for minimizing the quadratic function in equation \eqref{eq:quadEn}, where we choose $f(x)=s^T A_{\lin} s$ and $g(x)=s^T A_{\quadr} s + \Xi_{\{-1,1\}}(x)$, where $\Xi_{\{-1,1\}}(x)$ is the indicator function of the nonconvex set that restricts every entry of $x$ to be in $\{-1,1\}$. Consequently, we obtain 
\begin{align}
    s^{(k+1)}&= \argmin_{s\in\{-1,1\}^n} \quad  s^T A_{\quadr} s+ (s^{(k)})^T A_{\lin} s^{(k)} +  
    \innerp{ 2 A_{\lin} s^{(k)} }{ s-s^{(k)} } + \frac{\lambda}{2}\norm*{s-s^{(k)}}^2_2  \nonumber  \\
    &= \argmin_{s\in\{-1,1\}^n} \quad  s^T A_{\quadr} s +   
    \innerp{2  A_{\lin} s^{(k)} }{ s } - \lambda \innerp{ s}{s^{(k)}}. 
    \label{eq:Specific}
\end{align}
The above scheme is based on the fact that 
\begin{equation}
    \frac{\partial}{\partial s_i}(s^TAs)= (2As)_i
\end{equation}
holds for any symmetric matrix $A$.

The regularization term $\lambda\,||s-s^{(k-1)}||$ from the iteration step  
\eqref{eq:Specific} can be interpreted as a damping or regularization term to penalize strong deviations from the last iterate.
In the following, we present a weak monotonicity proof for a setting which is more general than in our equation \eqref{eq:Specific}. A proof like this can, for example, be found in Ref.~\cite{beck2009fastFirst}.

\subsubsection{Weak monotonicity proof} \label{sec:weak_mono}
\label{sec:Proof}
We consider a composite objective function of the form 
\begin{equation}
    E(x )= E_1(x) + E_2(x)\, ,
\end{equation}
which we want to optimize over a set $C\subseteq \mathbb{R}^n$.
For the individual parts, we assume that $E_1$ can be optimized directly while $E_2$ cannot, but we have access to its gradient $\nabla E_2$. To do this, we take the parts where the quadratic function exists and can calculate the gradient there analytically.
To solve the minimization problem
\begin{equation}
    \min_{x\in C} E(x ) \quad \text{for } E(x )= E_1(x) + E_2(x), 
\end{equation}
we linearize the $E_2$ part and solve the minimization iteratively. 
In more detail, we iterate in the same way as in equation \eqref{eq:ProxIt}:

\begin{align}
    x^{(k+1)}= \argmin _{x\in C} \quad   E^{(k)}(x)\coloneqq\ & E_1(x )+ E_2(x ^{(k)}) \nonumber \\&+ 
    \innerp{ \nabla E_2 (x^{(k)}) }{ x-x^{(k)} } + \frac{\lambda}{2}\norm*{x-x^{(k)}}^2 .  \label{eq:Abstract}
\end{align}

Obviously, $E^{(k)}(x^{(k)})= E(x^{(k)})$. 
Moreover, for any $x$ and sufficiently smooth $E$ we can apply 
the Taylor expansion of $-E_2$ with remainder estimation to obtain
\begin{align}
    E^{(k)} (x )-E(x)
&= 
    -E_2(x )+ E_2(x ^{(k)})  
    + \langle  \nabla E_2 (x^{(k)}) , x-x^{(k)}  \rangle+\frac{\lambda}{2}||x-x^{(k)}|| ^2 \nonumber \\
&= 
\bigl\langle  \frac{1}{2} \nabla^2 E_2 (\xi) (x-x^{(k)}), x-x^{(k)}  \bigr\rangle  +  \frac{\lambda}{2}||x-x^{(k)}||^2,  \label{eq:GeneralDerivation}
\end{align}
with a suitable $\xi\in C$.
In case $||\nabla ^2 E_2(\xi)||\leq c$ holds for a constant $c$ for all $\xi\in C$, we know that $-\frac{1}{2}\langle  \nabla^2 E_2(\xi)(x-x^{(k)}) , x-x^{(k)} \rangle \geq -\frac{c}{2} || x-x^{(k)}||^2$. 
Thus, we conclude that for a regularization parameter
$\lambda \geq c$
\begin{equation}
    E^{(k)}(x)\geq E(x) \quad  \forall x \in C\, .
\end{equation}
This bound implies that
\begin{equation}
    E(x^{(k+1)})\leq E^{(k)}(x^{(k+1)}) \leq E^{(k)} (x^{(k)})= E(x^{(k)}),
\end{equation}
and we have proven monotonicity.

\subsection{Statement for Ising problems}
\label{sec:forIsing}

We will now discuss some results that are specific to the case with discrete, binary variables. Note again that we have $E_2(s)= s^T A_{\lin} s $ in this case.
 For the regularization term we know that weak monotonicity will be guaranteed for a regularization parameter $\lambda$ greater or equal than the largest eigenvalue of $A_{\lin}$, since $||\nabla ^2 E_2 (\xi)||= || 2A_{\lin}||$, assuming $A_{\lin}$ is symmetric.  
The exact approximation that has been made after equation \eqref{eq:GeneralDerivation} can be described in the following way:
\begin{align}
    & s^T A_{\lin} s- (s^{(k)})^T A_{\lin} s^{(k)} - 2 s^{(k)} A_{\lin } (s - s^{(k)}) \nonumber \\
    &=  s^T A_{\lin} s+ (s^{(k)})^T A_{\lin} s^{(k)} - 2 s^{(k)} A_{\lin} s  \nonumber  \\
    &= (s- s^{(k )})^TA_{\lin}(s- s^{(k )})
\label{eq:SpecificQuadratic}
\end{align}
and the absolute value of the term was upper bounded as $(s- s^{(k )})^TA_{\lin}(s- s^{(k )})\leq \norm{A_{\lin}} \norm{s-s^{(k )}}^2$ to obtain an estimate for the regularization parameter $\lambda$.

One insight from equation \eqref{eq:SpecificQuadratic}
is that if $A_{\lin}$ has zeros on the diagonal, then for steps with exactly one changed entry, the approximation is perfect. 

Furthermore, since we are interested in the discrete case, it is also comparatively easy to provide a convergence analysis. Although this clearly does not hold for every interesting case, we assume here that every binary vector has a different energy, so that $ E(s_1) =s_1^T  A s_1+ b^T s_1 \neq  s_2^T  A s_2+ b^T s_2    =E(s_2)$ for $s_1 \neq s_2 $. From the previous analysis, it is clear that the sequence of energies $\left(E(s^{(k)})\right)_{k \in \mathbb{N} }$ is weakly monotonically decreasing. Since there are only finitely many possibilities for how the energy can decrease and one cannot go back to higher energies, the sequence of energies will converge. By excluding the case that two binary vectors have the same energy, the sequence of iterates also converges.

\subsection{Fixing variables, large neighborhood local search (LNLS)} \label{sec:FixVar}
A special case of the described method consists in fixing some variables and optimizing over the rest, as it is done in Raymond et al.~\cite{raymond2023hybrid} or in Refs.~\cite{beasley1998heuristic,rosenberg2016building} to solve large problems.
In each iteration $k$, we pick an index set $\mathcal{J}_k \subseteq\{1,2,\dots,n\}$ describing the variables we want to optimize. The other 
variables have per definition indices in $\mathcal{J}_k^c$ and are fixed to their previous value. 
Therefore, we consider the optimization problem
\begin{equation}
s^{(k)}= \argmin_{ \{  s \in \{ -1,1 \}^n \mid \forall j \in \mathcal{J}_k ^c:\ s_j= s^ {(k-1)}_j \} }  s^T A s
\end{equation}
to get the next iterate.
Explicitly writing down the sums yields
\begin{align}
s^{(k)}&= \argmin_{ \{ s \in \{-1,1 \}^n |  s \vert _{ \mathcal{J}_k^c  } = s ^{(k-1)}\vert _{ \mathcal{J}_k^c  } \} }  
 \sum_{ i \in \mathcal{J}_k } \sum_{ j  \in  \mathcal{J}_k  } (s_i) 
 ^T A_{i,j} s_j+ \nonumber \\
 &\sum_{ i \in \mathcal{J}_k } \sum_{ j  \in  \mathcal{J}_k ^c } \left( (s^{(k-1)}_i) 
 ^T A_{i,j} s_j + (s_j) 
 ^T A_{j,i }   s^{(k-1)}_i \right)  + \text{constant},
\end{align}
where $s\vert _{ \mathcal{I}  }$ indicates that one only looks at the vector elements for indices in the indexset $\mathcal{I}$.
For symmetric $A$ this reduces to
\begin{align}
s^{(k)}&= \argmin_{ \{ s \in \{-1,1 \}^n |  s \vert _{ \mathcal{J}_k^c  } = s ^{(k-1)} \vert _{ \mathcal{J}_k^c  }\} }  
 \sum_{ i \in \mathcal{J}_k } \sum_{ j  \in  \mathcal{J}_k  } (s_i) 
 ^T A_{i,j} s_j \nonumber \\ &+
 \sum_{ i \in \mathcal{J}_k } \sum_{ j  \in  \mathcal{J}_k ^c }  2 (s^{(k-1)}_i) 
 ^T A_{i,j} s_j  .
\end{align}
Note that this method is weakly monotonic, since we always optimize the original energy and just restrict the set of binary vectors that can be obtained for the next iterate.
The fact that we do not need damping for weak monotonicity here can be understood by looking at equation
\eqref{eq:SpecificQuadratic}. If we apply the absolute value on both sides of the equation, we see how much the linearization of $ s^TA_{\lin}s$ differs from the original term. 
Here, $A_{\lin}$ has a block with zeros (or can be permuted in such a way), and outside of this block variables do not change, so that $s_j= s_j ^{(k)}$. As a result the term on one side of the equation equals $0$.

Fixing a set of variables and optimizing over the remaining $m$ variables can also be called large neighborhood local search with $m$-flip neighborhood \cite{gonzalez2007handbook}. The $m$-flip neighborhood is the set of binary vectors, where not more than $m$ entries change with respect to the current iterate.
This is why we will abbreviate this method to 'LNLS' for large neighborhood local search, where $m$ is given in the numerical experiments.

\subsection{Relation to projected gradient descent}\label{sec:RelGradDesc}
Let us consider the special case with $A_{\quadr}=0$ and $A_{\lin}=A$. 
Now the iteration step becomes:
\begin{align}
    s^{(k+1)}
    &= \argmin_{s\in\{-1,1\}^n}     
    \innerp{2  A s^{(k)} }{ s } - \lambda \innerp{ s}{s^{(k)}}  \nonumber \\ 
    &= \argmin_{s\in\{-1,1\}^n}     
    -\innerp { s }{  s^{(k)}-2  \frac{1}{\lambda} A s^{(k)}}.       \label{eq:OnlyLin}
\end{align}
The $\argmin$ together with the scalar product with $-s$ can be seen as taking the sign of every component of $\lambda s^{(k)}-2  A s^{(k)}$. Furthermore, one can describe this as a projection onto the vertices of the hypercube $\{-1,1\}^n$. The term that is projected is the last iterate minus the gradient of the function we want to optimize times a constant. Therefore, we have found a relation to the projected gradient descent formula. However, it has to be noted that we do not operate on a convex set, but instead on individual points. This restriction makes it difficult to apply various results about the projected gradient descent algorithm to our case. The relation from ISTA to projected gradient descent holds even in the general case, as can be seen in Ref.~\cite{beck2009fast}.

\subsubsection{Multiple relaxations leading to different gradients.}
Notice that the diagonal entries of the matrix $A$ in equation \eqref{eq:Ising} do not matter for the optimization problem if we optimize over $-1$ and $1$ as entries.
However, for the linearized term $\innerp{2A_{\lin}s^{k}}{s}$ they have an effect in places where $s^{(k)}_i \neq s_i$. If the last two terms of equation \eqref{eq:Specific} were separate, this would lead to projected gradient descent. We will now further investigate the effect of the diagonal elements.
The objective function 
\begin{equation}
    f(s)= s^TAs + b^T s
\end{equation}
with variables $s_i\in \{-1,1\}$ can have multiple possible relaxations to variables $s_i \in [-1,1]$. We can see this by adding and subtracting a sum of arbitrary constants $c_i$:
\begin{align}
    f(s)&= s^TAs + b^T s + \sum_i c_i- \sum_i c_i  \nonumber  \\
      &= s^TAs + b^T s + \sum_i c_i s_i^2- \sum_i c_i.
\end{align}
Inserting the $s_i^2$ as factor does not lead to changes for $s_i\in \{-1,1\}$ but yields a relaxation with different values in $s_i\in (-1,1)$. Considering a relaxation also allows us to look at gradients. For the gradient we have
\begin{align}
    \nabla f(s) = 2s^T A  +b + 2c \odot s
\end{align}
with $c= (c_1,..., c_n)$ and ``$\odot$'' denoting element-wise multiplication. 
Since we want to treat every variable in the same way, we set $c_1=c_2=...=c_n$. 
However, we can imagine that there is some better way to make use of the $c_i$ parameters.

\subsection{Permutating the adjacency matrix of the hardware graph} \label{sec:Permutating}
\begin{figure}
    \centering
    \includegraphics[width=\linewidth]{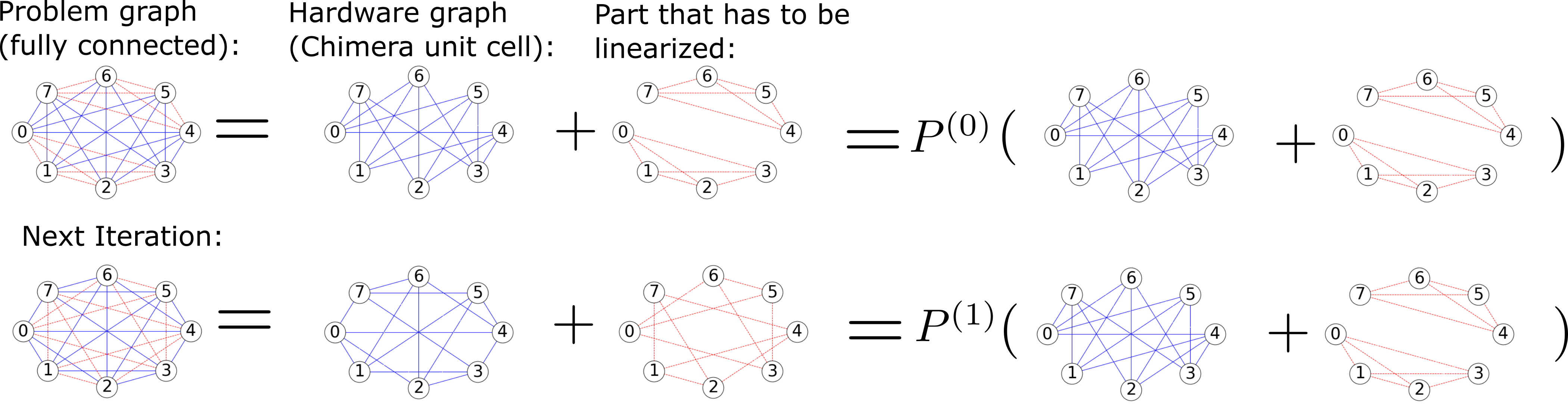}
    \caption{The first row shows the decomposition of a fully connected problem into a part that fits the hardware graph of a Chimera graph unit cell and a part of the edges that do not fit. In the second row, 
   a random permutation is applied to the hardware graph to get a different decomposition. The edges without resemblance in the hardware graph influence only the biases in our method. Because of the random permutation, the part that needs to be approximated changes in every iteration. 
    }
    \label{fig:Permutation}
\end{figure}

Instead of always linearizing the same parts of $A$, we can permute the adjacency matrix $M$ of the hardware graph in each iteration $k$ with a random permutation $P^{(k)}$. The principle is visualized in \cref{fig:Permutation}.
Applying this permutation is important because the linearization is an approximation, and varying the subproblems we approximate leads to better results.
Therefore, we extend the iteration step from \cref{eq:Specific} to:
\begin{align}
A_{\quadr}^{(k)}:&= ((P^{(k)})^T M  P^{(k)}) \odot A    \\
A_{\lin}^{(k)}:&= A- A_{\quadr}^{(k)}    \\
   s^{(k+1)} &= \argmin_{s\in\{-1,1\}^n}   s^T A_{\quadr}^{(k)} s  +  
    \innerp{2  A_{\lin}^{(k)}  s^{(k)} }{ s } - \lambda \innerp{ s}{s^{(k)}}. \label{eq:SpecificPerm}
\end{align}
The whole process can be found in \cref{alg:cap}.
The regularization parameter $\lambda$ can be chosen as described in \cref{sec:Proof}. However, in experiments, we often encounter situations where the method is stagnant. To improve results, we allow for $\lambda$ parameters smaller than $||A_{\lin}||$ from \cref{sec:Proof}.
The next section goes into detail about how to choose the regularization parameter $\lambda$.

\begin{figure}
    \centering
    \includegraphics[width=0.9\linewidth]{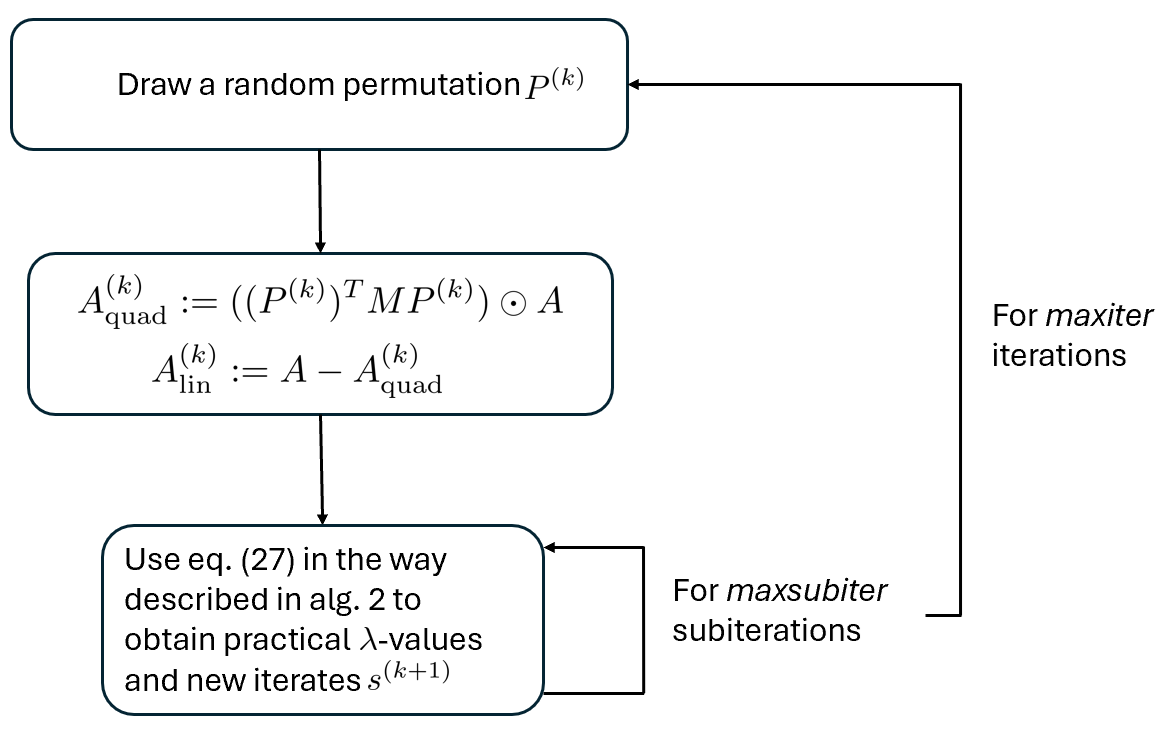}
    \caption{ 
    Workflow diagram of our method. 
    In the iterations, we change the permutations, while the $\lambda$ parameters are dealt with in the subiterations.}\label{fig:workflow}
\end{figure}
\begin{algorithm}
\caption{Drawing a new permutation every time}\label{alg:cap}
\begin{algorithmic}[1]
\Require Adjacency matrix of hardware graph $M\in \{0,1\}^{n \times n}$ and coupling matrix $A\in \mathbb{R}^{n \times n}$, 
Solver: $S(B,c)= \argmin_{x\in \{ -1,1\}^n}  x^T Bx+ c^Tx $ for $B\in \mathbb{R}^{n \times n}$ with $B_{i,j}=0$ if $M_{i,j}=0$
 \Ensure Try to minimize $x^TAx$
\State Randomly initialize $x_\textrm{A}$
  \For{$i\gets0$ \textbf{to}
  \text{maxiter} }
  \State 
$ x_{\textrm{Aold}}= x_{A} $ \label{line:FirstAssignment} 
\State  Sample a random permutation $P_k$
  \State  $A_{\quadr } = M\odot (P_k^T A P_k )$
\State   $ A_{\lin}  = P_k^T AP_k-  A_{\quadr }$
  \State    \label{line:FinalStep} 
  $ x_\textrm{A}  = P_k S(A_{\quadr},  P_k^T  ( 
 \left( P_k A_{\lin} P_k^T \right)  x_{\textrm{Aold}}- \lambda x_{\textrm{Aold}} )) $ 
   \EndFor
\end{algorithmic}
\end{algorithm}

\subsection{Strategy for regularization parameters}
\label{sec:Strategies}

To get practical 
values for the regularization parameter $\lambda$, we have a closer look at the linear term only:
\begin{equation}
  L:=   P_k^T  ( 
 \left( P_k A_{\lin} P_k^T \right)  x_{\textrm{Aold}}- \lambda x_{\textrm{Aold}}.
\end{equation}
In the special case $A= A_{\lin}$, the sign of the entries in this vector determines if the iterate has changed entries.
The idea for determining $\lambda$ values is to order the entries of $\abs{L}$ by size and denote this as $|L|_{ \ord}$. Then, the values we try out for lambda are restricted to:
\begin{equation}
    \lambda \in S_{\lambda}\subseteq S_{\textnormal{sign-change}}:= \Bigl\{v\in \mathbb{R} \mid  \exists i \in \{0,..., n \} : v= \frac{(|L|_{ \ord})_i +(|L|_{ \ord})_{i+1}}{2}   \Bigr\}. \label{eq:Slambda}
\end{equation}
The goal of this construction is to avoid trying out multiple values for the regularization parameter $\lambda$ that result in the same next iterate.
The process for regularization parameter selection can be seen in \cref{alg:Increasing}. The algorithm is meant to be executed in \cref{alg:cap} as replacement for line \ref{line:FinalStep}
and line \ref{line:FirstAssignment} is removed. In \cref{fig:workflow} there is a workflow diagram of the method, showing the nested for-loops.

\begin{algorithm} 
\begin{algorithmic}[1]
\Require{ Adjacency matrix of hardware graph $M\in \{0,1\}^{n \times n}$, coupling matrix $A\in \mathbb{R}^{n \times n}$ and current permutation $P$}
 \Ensure{ Determine good $\lambda$ parameters} 
  \State Compute $ S_{\textnormal{sign-change}}$ as in \eqref{eq:Slambda}  \\
   $S_\lambda$ contains every $|S_{\textnormal{sign-change}}|/\text{maxsubiter}$-th element from $ S_{\textnormal{sign-change}}$ ordered according to size
   \State $\text{energyBest}= E(x_{\textrm{Aold}})$
   \State $x_{best}=x_{\textrm{Aold}}$  
  \For{$i\gets0$ \textbf{to} \text{maxsubiter} }
    \State $\lambda= \left( S_\lambda\right)_i $
    \State $x_{\textrm{Anew}}= P_k S(A_{\quadr},  P_k^T  ( 
 \left( P_k A_{\lin} P_k^T \right)  x_{\textrm{Aold}}- \lambda   x_{\textrm{Aold}} ))$ \If{$\textnormal{energyBest}\geq E(x_{\textrm{Anew}})$} 
    \State $x_{best}= x_{\textrm{Anew}}$
    \State $\text{energyBest}= E(x_{\textrm{Anew}})$
    \EndIf
    \EndFor
   \State $x_{\textrm{Aold}}= x_{best}$
 \caption{How to deal with $\lambda$ parameters. }\label{alg:Increasing}

 \end{algorithmic}
\end{algorithm}

\subsection{Choosing a part of the hardware graph with high connectivity}\label{sec:HighConnectivity}
\begin{figure}
\centering
\subfloat[ ]{ \includegraphics[trim={2cm 0cm 2cm 0cm},clip,width=0.45\linewidth]{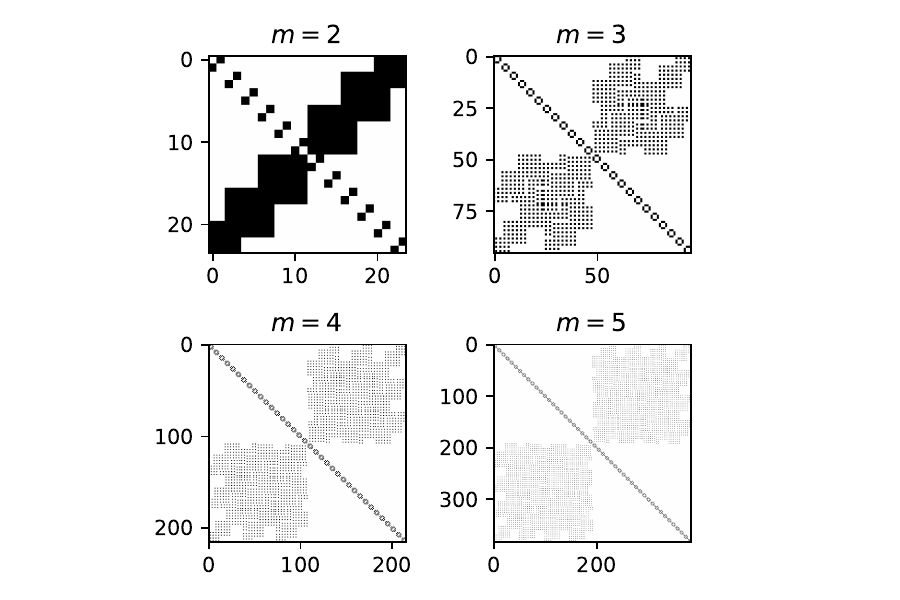} }
\subfloat[ ]{ \includegraphics[trim={2cm 0cm 2cm 0cm},clip,width=0.45\linewidth]{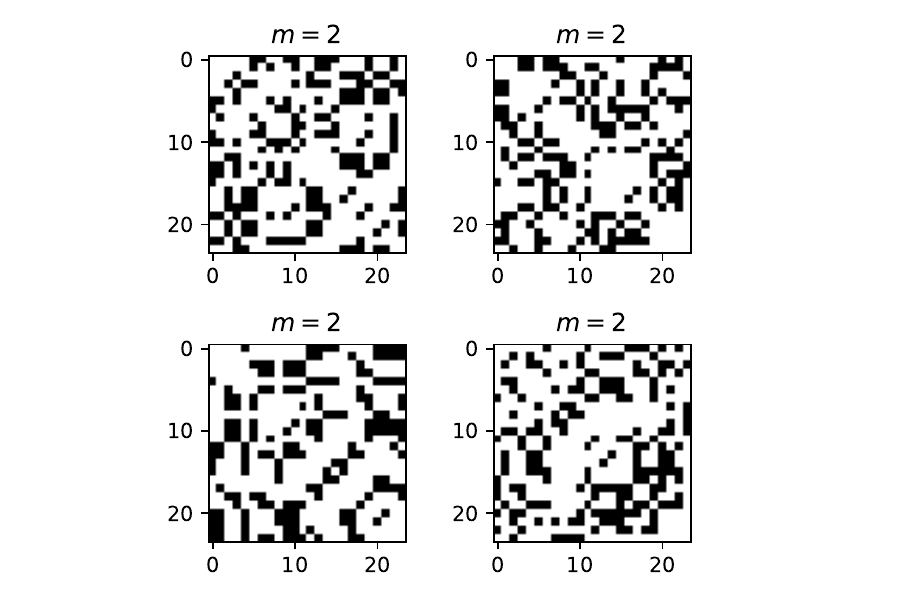} }
    \caption{(a) Adjacency matrices for different sizes of the Pegasus graph. (b) Adjacency matrices for a Pegasus graph of size $2$ but randomly permuted. }
    \label{fig:Adj}\label{fig:AdjPerm}
\end{figure}

If the problem matrix $A$ is smaller than the full hardware graph, we have to consider a subset of the qubits. This is done considering only qubits that belong to a Pegasus graph with a smaller size parameter $n$ \cite{boothby2020next}.
In \cref{fig:Adj}, the adjacency matrices for Pegasus graphs with different sizes and what random permutations of this look like can be seen.

\subsection{Simulated annealing-like regularization} \label{sec:SAlikeReg}
 While monotonicity is a very important property for the theoretical investigation of an optimization algorithm, it has the drawback that one can get stuck in local minima. How often or if this occurs is, of course, dependent on the neighborhood structure and the problem instances themselves. Therefore, in many settings, it is often beneficial to allow steps where the energy does not decrease to escape local optima with some probability.
Simulated annealing, in particular, is designed so that it allows steps that do not improve the energy. 
In the following, we show that it is possible to formulate simulated annealing as a series of simple binary, linear optimization problems.
We will use the same idea to break the monotonicity of optimizing random subproblems in an attempt to make it more likely to get to the global optimum. 
Let $s^{(k-1)}$ be the current iterate.
A random neighbor of $s^{(k-1)}$ is constructed by first drawing a random index set $\mathcal{J}_k$ and then flipping all values that correspond to indices in $\mathcal{J}_k$.
This vector we name $s_N^{(k)}$.
Now the next iterate is constructed with
\begin{equation}
    s^{(k)}= \argmin_{ \{ s \in \{-1,1 \}^n |  s \vert _{ \mathcal{J}_k^c  } = s ^{(k-1)}\vert _{ \mathcal{J}_k^c} \} }     \sum_{ i \in \mathcal{J}_k}
     (  e^ { -\frac{E(s_N^{(k)})-E(s^{(k-1)})}{T(k)} } -R_k)s_i s^{(k-1)}_i,
\end{equation}
where $R_k \in  \left[0,1 \right] $ is for each iteration uniformly random generated and $T(k)$ is a typical temperature schedule for simulated annealing.
An example for such a schedule would be $T(k)= \frac{c}{\log(1+k)}$ with a fixed constant $c\in \mathbb{R}$.
The idea is to combine this with our QUBO solver like this:
\begin{equation}
        s^{(k)}= \argmin_{ \{ s \in \{-1,1 \}^n |  s \vert _{ \mathcal{J}_k^c  } = s ^{(k-1)}\vert _{ \mathcal{J}_k^c }\} }  s^T A s +   \sum_{ i \in \mathcal{J}_k}
     (  e^ { -\frac{E(s_N^{(k)})-E(s^{(k-1)})}{T(k)} }-R_k )s_i s^{(k-1)}_i. \label{eq:simpleComb}
\end{equation}
One can even add factors to both summands to weight the contributions of the simulated annealing and the non-linear QUBO term. In principle, this should yield 
an optimization algorithm similar to the combination
of simulated annealing and local search proposed in Ref.~\cite{martin1996combining}. The simulated annealing part of the algorithm can only jump to vectors that
have optimal energy for a fixed Hamming distance to the previous iterate. We leave the numerical verification of this idea to future research since there is still some thought required in finding the best weighting factors and a good schedule. However, we still present it here since it fits the discussion of damping terms in these iterative algorithms.

\begin{figure}
\subfloat[ ]{\includegraphics[scale=0.4]{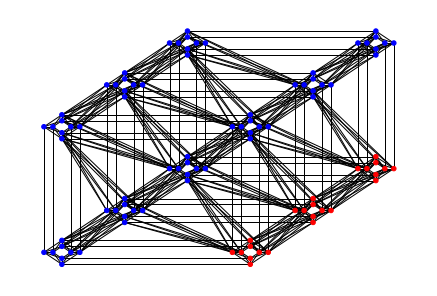}}
\subfloat[ ]{ \includegraphics[scale=0.4]{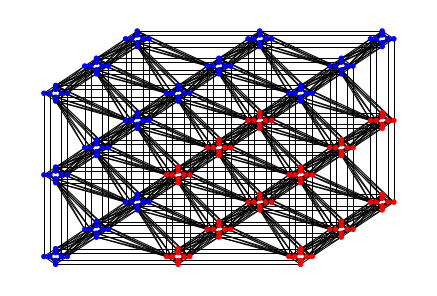} }
    \caption{The smaller Pegasus sub-graphs indicated by the red vertices in (a) and (b), respectively, are part of the bigger Pegasus graphs that also include blue vertices.}
    \label{fig:PGraph}
\end{figure}

\section{Results}\label{sec:Experiments}
In the following, we demonstrate the performance of our method by using simulated annealing as a QUBO sampler and artificially restricting it to the D-Wave Pegasus Hardware Graph. For the experiments with simulated annealing, we used the D-Wave neal solver \cite{DWave_Neal} with 100 runs per problem. 
We also demonstrate the functionality of our method on the D-Wave Advantage machine version 6.4. All the methods that were tested and the section or reference where they were described can be found in \cref{tab:my_table}.

\begin{table}
    \centering

    \begin{tabular}{|c|c|c|} \hline 
       Abbreviation  & Name & Reference \\ \hline \hline
       Splitting  & \makecell{Here presented method based on \\ proximal gradient descent}   & s. \cref{sec:LiDa,sec:Permutating,sec:Strategies}\\   \hline 
      LNLS   & Large Neighborhood Local Search & s. \cref{sec:FixVar} \\ \hline 
      SA on full problem &\makecell{ Application of the simulated \\ annealing solver Neal } &  \cite{DWave_Neal}\\ \hline
      SA hardware-restricted &\makecell{ Neal solver artificially restricted \\ to D-Wave Pegasus topology} &  \cite{DWave_Neal}\\
      \hline
      k-Opt &k-Opt with k-swap neighborhood  &  \cite{merz2002greedy}\\
      \hline
    \end{tabular}
\caption{Abbreviations and references for all the tested methods.}
\label{tab:my_table}
    
\end{table}

\subsection{Datasets}
As problem instances, we investigate the 
regular spinglass instances from \cite{mehta2022hardness} and part of the Max-cut problems from MQLib \cite{DunningEtAl2018}, which we formulate as QUBOs.

In Ref.~\cite{mehta2022hardness} among other QUBOs, a fully-connected regular spin-glass model is introduced for benchmarking. The publication also provides comparisons between the hybrid-quantum method qbsolv \cite{qbsolv}, a direct solution with the D-Wave Advantage using minor embeddings, and a baseline exhaustive search. They also experiment with Ising problems with randomly drawn coefficients. This is similar to the well-known Sherrington-Kirkpatrick model \cite{sherrington1975solvable} but with uniform probability distributions instead of a Gaussian distribution.

The regular spin-glass model, Ising problems have the form
\begin{align}
    A_{i,j}&= 1-\frac{i+j-2}{N-1}, \textnormal{ for } i \neq j    \\
    b_i &= 1- 2 \frac{i-1}{N-1},
\end{align}
for a fixed dimension $N$.
We abbreviate these instances with "Reg".
\\\\

 In a Max-Cut problem, one seeks to find a cut through a weighted graph so that the sum of the weights of the edges that are cut is as large as possible. Max-Cut optimization problems are fully equivalent to QUBOs. However, for the formulation of an arbitrary QUBO problem with $n$ variables, a Max-Cut problem with $n+1$ vertices is needed \cite{QUBOMAXCUT}. The problem instances in the MQlib dataset range from applications in very large-scale integrated circuit design and telecommunication networks, to integrative biological network analysis or image segmentation \cite{DunningEtAl2018}. Multiple instances are also generated using graph theoretical, mathematical models, for example, via Watz--Strogatz graphs.

\subsection{Regular spinglas dataset from Ref.~\cite{mehta2022hardness}}

We will benchmark the methods presented here on the regular spinglas problems from Mehta et al.~\cite{mehta2022hardness}.
For the subproblems, a simulated annealing solver is used; for our splitting method, the connectivity is restricted to the Pegasus Hardware graph.

\begin{figure}[!tbp]
  \subfloat[ ]{\includegraphics[trim={2cm 0 1cm 0},width=0.5\textwidth]{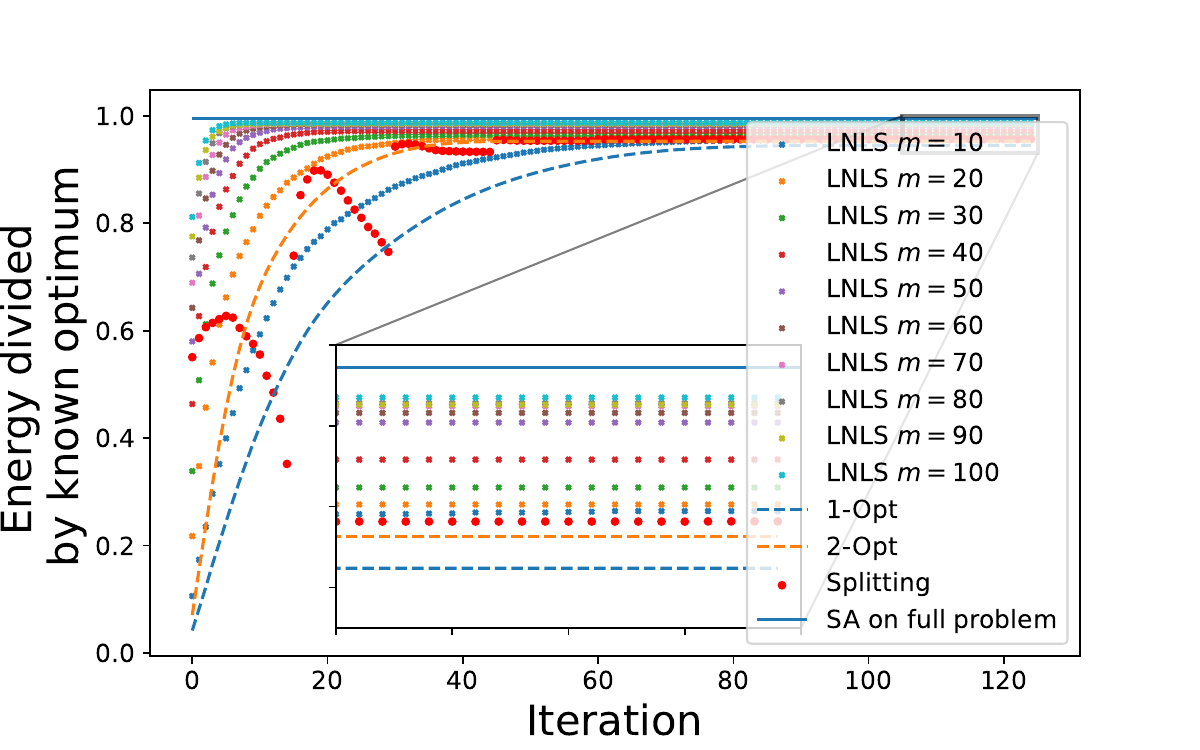}}
  \hfill 
  \subfloat[]{\includegraphics[trim={1cm 0 2cm 0},width=0.5\textwidth]{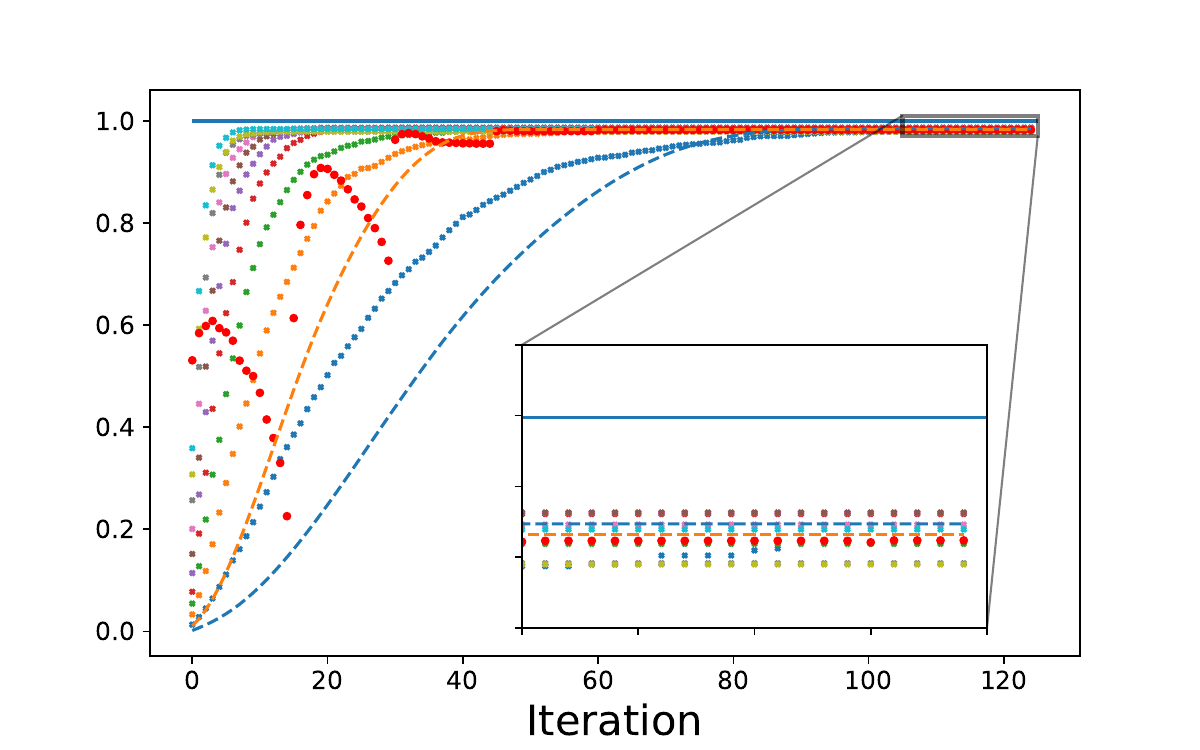}}
  \caption{ Approximation ratio depending on the iteration, e.g. number of neal QUBO-solver calls, averaged over (a) all Reg instances, (b) all Reg instances greater than $150$. 
  The plot shows how the optimization progresses for the different methods.
  } 
  \label{fig:RegVerlauf}
\end{figure}

\Cref{fig:RegVerlauf} depicts how the averaged approximation ratio $\frac{E}{E_{\textrm{opt}}}$ decreases over the iterations.
One iteration is one call of the simulated annealing QUBO solver
Neal
 \cite{DWave_Neal}. Therefore, the subiterations from \cref{alg:Increasing} are also visible in the plot. For all the reported experiments, we have used $15$ as the maximal number of subiterations.

For the calculation of the approximation ratio, the ground state needs to be known. 
Here we use a conjecture from Ref.~\cite{mehta2022hardness} about the ground state, which we prove in appendix \cref{secA1}. The conjecture states that the optimal vector for the particular instances has all the negative entries at the beginning, and after the negative entries, there are only positive entries. Knowing this, one can easily compute the exact ground state by trying out all $n+1$ possibilities. The term splitting in the legend of \ref{fig:RegVerlauf} refers to our new proposed method. 1-Opt and 2-Opt refer to iteratively optimizing brute force in a 1 or 2 swap neighborhood, as we already mentioned in the related work \cref{sec:Rel}.

\begin{figure}[!tbp]
  \centering
  \subfloat[ ]{\includegraphics[trim={2cm 0 1cm 0},width=0.5\textwidth]{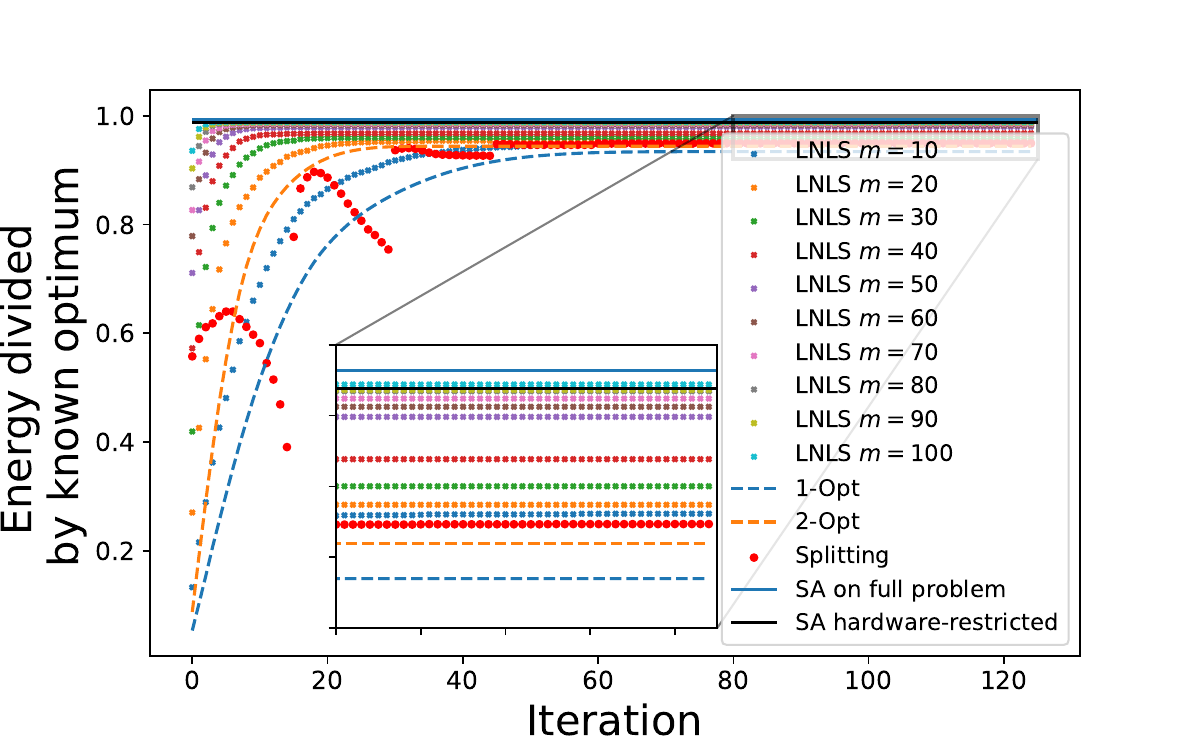}}
  \hfill 
  \subfloat[]{\includegraphics[trim={1cm 0 2cm 0},width=0.5\textwidth]{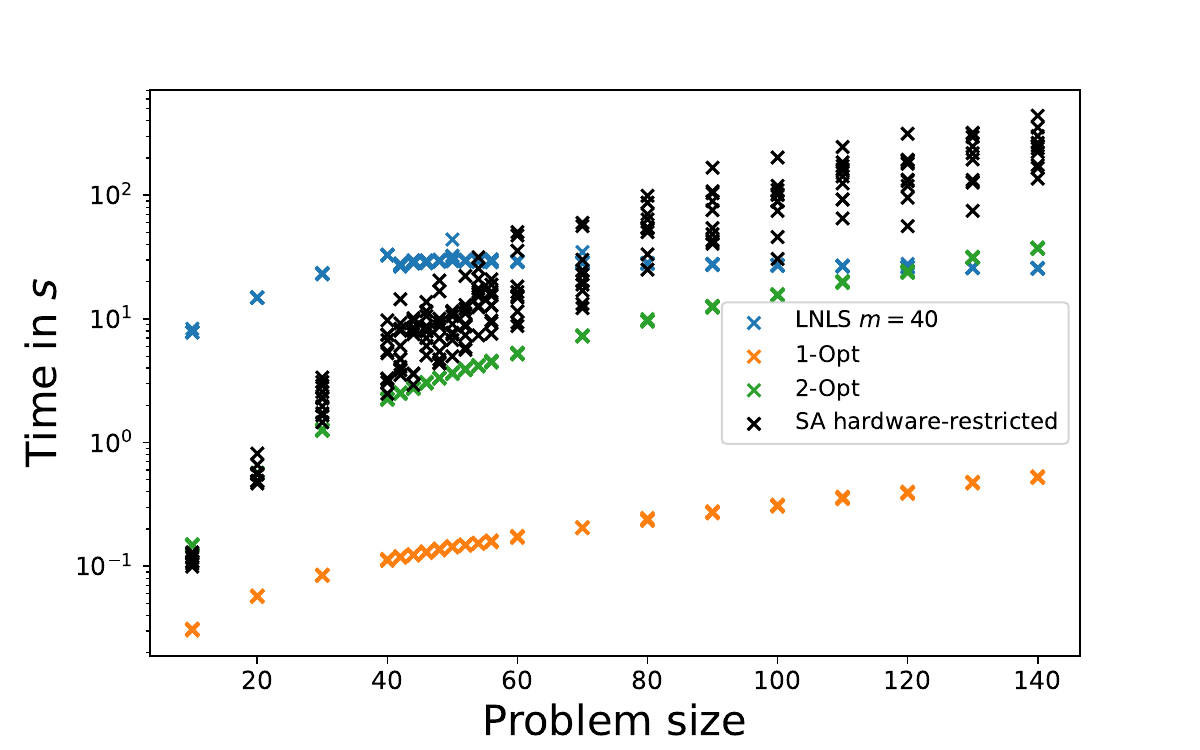}}
  \caption{ (a) Approximation ratio for all instances where an embedding could be found. An embedding was found in
$210$ from the $270$ Reg instances. The time limit for finding an embedding was $10$min. (b) Wall clock time consumption for the different methods on the Reg instances. The method where the minor embedding is computed can be found as ``SA hardware-restricted" in the legend. Although the ``SA hardware-restricted'' method outputs almost the best energies, it also has the worst time consumption for the bigger instances.}
  \label{fig:RegVerlaufMinorEmb}
\end{figure}

The experiments show that our method can handle bigger instances, which do not fit on the hardware graph completely, much better. 
Restricting the number of instances to those that are bigger than 150, as is done in 
\cref{fig:RegVerlauf} (b), improves the performance of the splitting compared 
to plainly solving subproblems. The same statement can be made for the k-Opt methods. Note that the smallest size of the subproblems is 10, which is already the size of the smallest instances. We also used a simulated annealing solver that is artificially restricted to the hardware connectivity and obtained solutions via minor embedding. This is named ``SA hardware-restricted'' in the plots. In \cref{fig:RegVerlaufMinorEmb} (a), we plot the behavior for all instances where an embedding is found. Note that although the Reg instances are fully connected, we did not use pre-computed embeddings but the default minorminer code. 
The time consumption of calculating minor embeddings can easily exceed the time the other methods take. This is depicted in the (b) part of \cref{fig:RegVerlaufMinorEmb}. 
\\

\subsection{Maximum cut problems} \label{sec:ExperimentsMQlib}
We test our methods on some of the instances from MQLib \cite{DunningEtAl2018}. 
While the MQLib contains in total $3396$ instances, we only look at 909 instances with problem sizes between $500$ and $1000$. 
The maximum cut problem is equivalent to a QUBO, as can be seen, for example, in Ref.~\cite{QUBOMAXCUT}. 
We compare the performance on these instances for varying dimensionality $m$ of the LNLS (s. \cref{sub:sizes}). In \cref{sub:dwave} we also show results from runs on the D-Wave Advantage quantum annealer. For the other experiments, we always used simulated annealing solvers. 
In the appendix \cref{sub:ablation}, we run an ablation study on this dataset to compare with a method that randomly applies permutations to the hardware graph without taking previous iterates into account and then outputs the best result.

\subsubsection{Comparison with various subproblem sizes} \label{sub:sizes}
\begin{figure}
    \centering
    \includegraphics[trim={2cm 0cm 2cm 0cm},width= 0.5\textwidth]{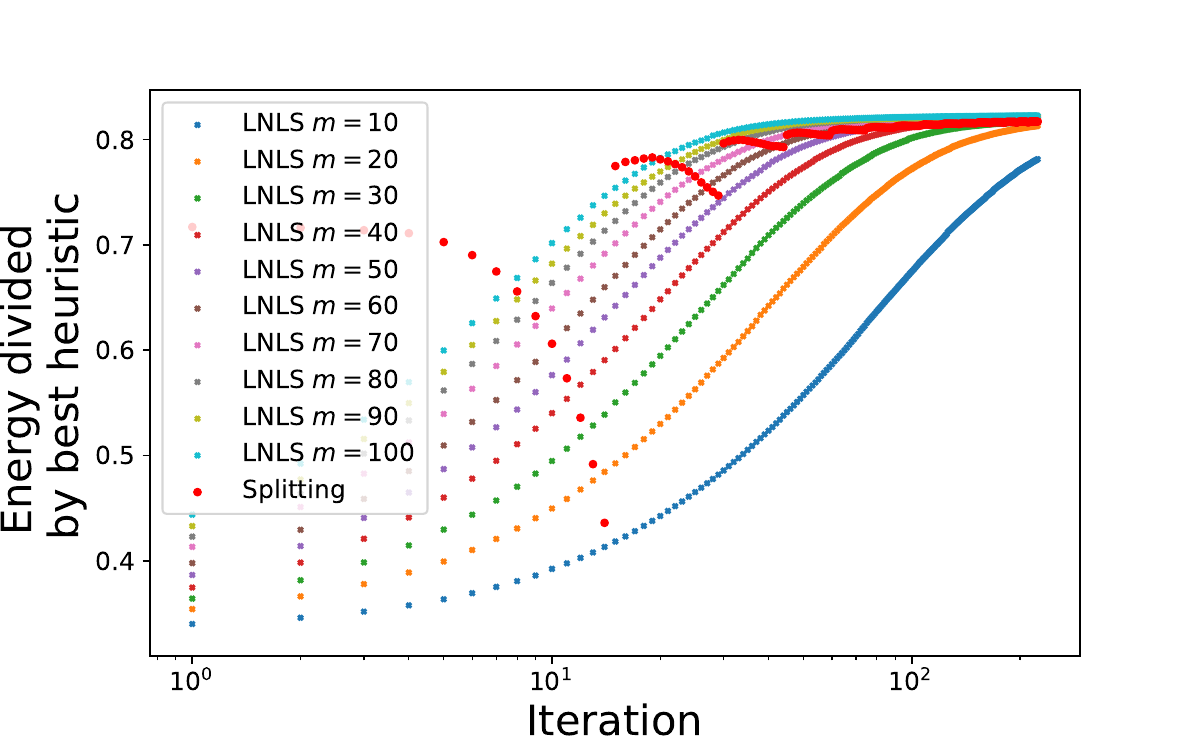}
    \caption{ Approximation factor averaged over 909 Instances from MQlib depending on the iteration. Which method performs best largely depends on the size of the subproblems. Bigger subproblems lead to faster improvement in the energy on average.} 
    \label{fig: MainExpVerlauf}
\end{figure}

We tested our splitting method as well as the plain calculation of subproblems with different sizes on the 922 problems from MQlib that have sizes between 500 and 1000 vertices. Results of this can be seen in \cref{fig: MainExpVerlauf}. In the approximation factor plot, we divide the absolute energy that is obtained by the heuristic with the best reported energy in Ref.~\cite{DunningEtAl2018}. The $13$ instances where the best reported energy is 0 are not used for this plot.
In the last iteration, the splitting method performed on par with optimization over subproblems of size 30. Calculating bigger subproblems achieved better energy, while using smaller subproblems achieved worse energies.

Note that the problems are already so big that one cannot find a minor embedding to the D-Wave Advantage machine. 
The 2-Opt optimization problem also does not finish in a reasonable amount of time for the big instances. That computation time is a bottleneck here for 2-Opt is to be expected since the time scales with $\mathcal{O}( \frac{n^2}{2}  )$.
Nevertheless, we also computed the 2-Opt method for completeness, and the results can be seen in the appendix \cref{sec:kOptExp} in  \cref{fig:MQLibwithkOpt}.

\begin{figure}[!tbp]
  \centering
  \subfloat[ ]{\includegraphics[trim={2cm 0 0.5cm 0},width=0.5\textwidth]{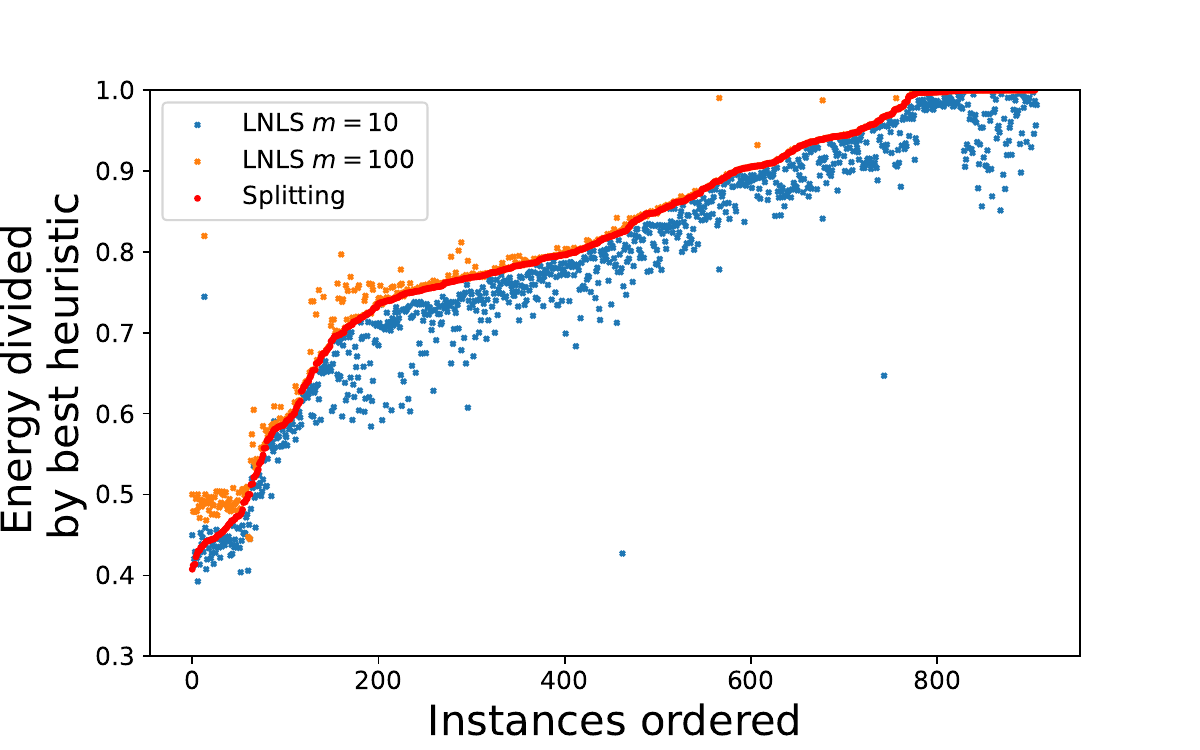}}
  \hfill 
  \subfloat[]{\includegraphics[trim={0.5cm 0 2cm 0},width=0.5\textwidth]{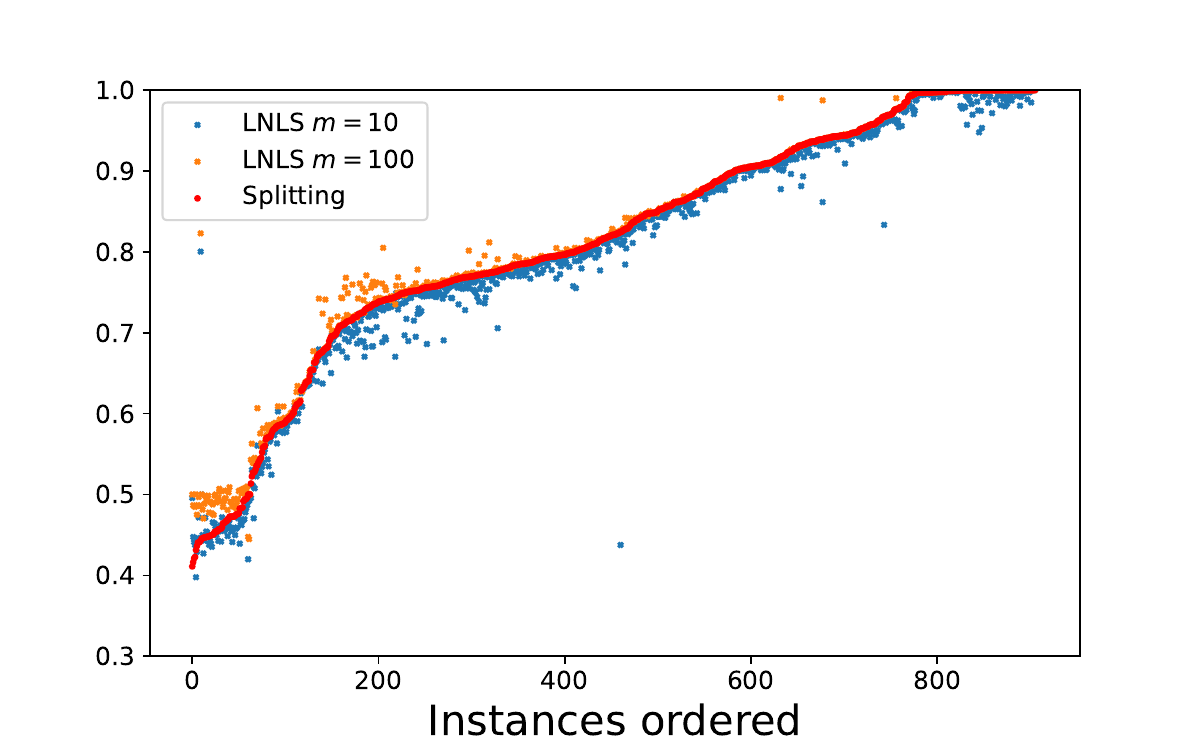}}
  \caption{Results for all the instances compared to the best heuristic for (a) 15 and (b) 25 iterations. The instances are ordered according to the performance of our splitting method. It can be observed that the performance compared to the best heuristic is largely correlated among the methods.} 
  \label{fig:AllInstances}
\end{figure}

\begin{figure}[!tbp]
  \centering
  \subfloat[ ]{\includegraphics[trim={2cm 0 0.5cm 0},width=0.5\textwidth]{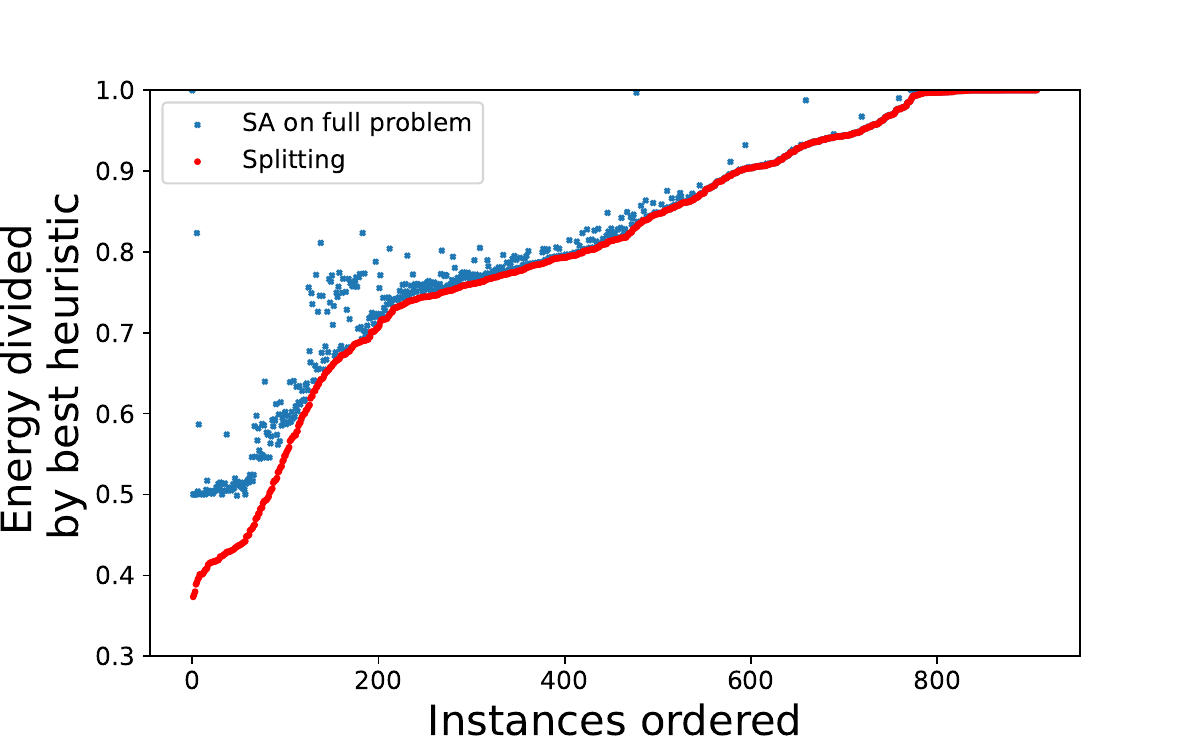}}
  \hfill 
  \subfloat[]{\includegraphics[trim={0.5cm 0 2cm 0},width=0.5\textwidth]{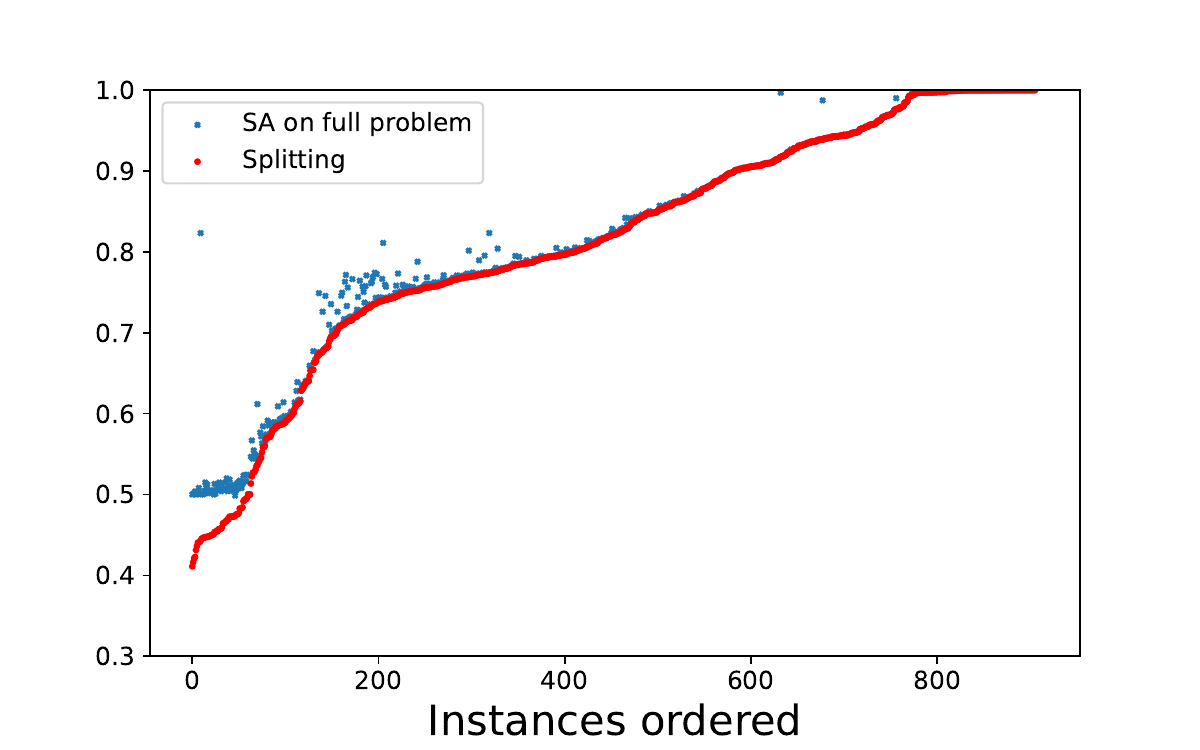}}
  \caption{Results for all the instances compared to the best heuristic for (a) 3 and (b) 25 iterations of the splitting method. The instances are ordered according to the performance of our splitting method. The performance of simulated annealing without hardware restrictions is also correlated to the performance of our splitting method compared to the best heuristic. Note that the values presented for the unrestricted simulated annealing are the same on both plots but can be reordered according to the performance of the splitting method.   }
  \label{fig:AllInstancesSA}
\end{figure}

Furthermore, we present the final energy of each instance divided by the best heuristic for each instance in \cref{fig:AllInstances,fig:AllInstancesSA}. For this depiction, the instances were ordered according to the performance of our splitting method.\\
It can be observed that a good performance compared to the best heuristics is strongly correlated among the different methods. Further analysis revealed that the instances where LNLS or our splitting method performed the best are often given by denser graphs. It also appears to be beneficial to have edge weights that are restricted to attaining only a few discrete values. Among the instances where the here presented splitting method coincides with the best heuristic were multiple Barabasi Albert \cite{barabasi1999emergence} and Powerlaw cluster \cite{holme2002growing} random graphs. On one really sparse and two very dense instances, our method performed better than the best heuristics that we found in the MQLib repository \cite{DunningEtAl2018}. But since for these three instances, even LNLS with $m=10$ achieved better results, we do not want to focus on this point. The instances are named ``g000098", ``g002324" and ``g002064". 
For the instances where our method performed worst compared to other heuristics, we observed a trend that those instances have rather sparse graphs. We assume that in these cases, other heuristics can better make use of the specific, sparse structure. A lot of the difficult instances are described to be models for solving real-world Very-Large-Scale Integrated Circuit \cite{koch1998solving} problem. The instance where our method performed worst was a hypercube graph \cite{resende2004new,biazzo2012performance}.

\subsubsection{Implementation on D-Wave Advantage} \label{sub:dwave}
We tested our methods on the D-Wave Advantage machine with version 6.4. 
Throughout the experiments, we used the default schedule and the default annealing time of $20\mu s$. The chain strength is chosen to be of factor $1.001$ stronger than the biggest absolute value in the couplings or biases. From the obtained state histogram, we only work with the state of lowest energy that has been obtained. Note that while even stronger chain strengths can reduce the chain break frequency for some problems, they also lead to a reduced effective precision, which can impair performance. If chain breaks occur, they are dealt with by using the default majority voting strategy. Overall, we have observed that in our benchmark, a chain strength of $1.001$ is beneficial over significantly larger values.

\begin{figure}[H]
    \centering
    \includegraphics[width=0.7\textwidth]{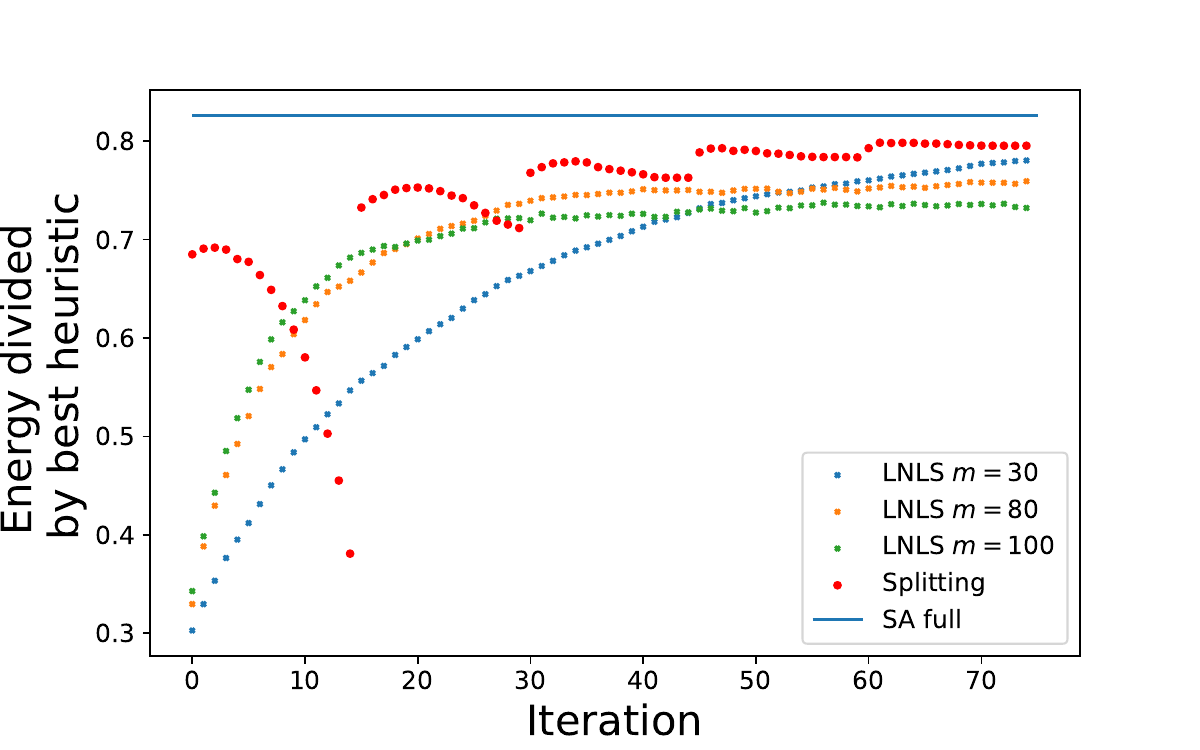}
    \caption{Energies divided by the best reported heuristic averaged for 13 Instances of MQlib that have $-1,1$ as coefficients. The experiments were carried out on the D-Wave Advantage quantum annealer except for ``SA full''. Therefore, we have provided a proof of concept for our iterative splitting method.
    }
    \label{fig:QuantumMain}
\end{figure}

By trying out different instances, we came to the conclusion that it is important what degree of precision is required for the QUBO matrix. This is well known, and methods to make a QUBO problem better suited for 
precision-restricted solvers like the D-Wave machines with experimental errors in the couplings have already been developed \cite{muecke2023optimumpreserving}.
We observed failure cases where our method never improved the energy or even made it worse. For example, it failed to optimize the instance ``imgseg\_271031'', which is motivated by an image segmentation problem and has float-valued weights in the graph. But also, the instance ``gka.5f'' with integer values between -50 and 50 was not optimized properly. 
At this point, to circumvent the problem, we tried out several instances with $-1,1$  as possible values for the edges and could confirm that our method works on the quantum annealer. The energy decreased in the same way as we observed in the simulated annealing experiments. In \cref{fig:QuantumMain}, the average over 13 instances of this sort is depicted.

\section{Conclusion}\label{sec4}
The splitting techniques from our method were, for the first time, investigated for discrete QUBO optimization. We obtained the theoretical insight that this method can be constructed weakly monotonically. Nevertheless, for the actual experiments, the method was more practical in a non-monotonically decreasing setting to avoid stagnant behavior. The comparison with the standard large neighborhood local search in this setting showed that the better performance is mostly dependent on the size of the local neighborhood. Most of our benchmark instances were too large for a direct minor embedding of the problem, but it is well known that these methods can take a long time and can lead to large chains that decrease performance. This was also confirmed for the part of the regular spin glass instances that was small enough. Whether calculating minor embeddings or using iterative algorithms is the better strategy is, of course, highly dependent on the problem, and we cannot give a final answer to this. The question of which characteristics of the problem instances can predict whether the method with splitting performs better than the other iterative methods is also largely unanswered. However, in the experiments using the actual quantum device, we obtained promising results on some Max-Cut problems that have $+1$ and $-1$ as coefficients. In general, we hope this work  encourages further research in splitting methods for QUBO optimization via hardware-restricted solvers. We will publish our source code at the  repository \cite{MyGithub}.

\begin{appendices}

\section{Proof for the conjecture about regular spin glasses from Ref.~\cite{mehta2022hardness}}\label{secA1}
The problem instances are constructed according to
\begin{align}
    A_{i,j}&= 1-\frac{i+j-2}{N-1}, \textnormal{ for } i \neq j  \label{eq:CouplingMehta}  \\
    b_i &= 1- 2 \frac{i-1}{N-1}.
\end{align}
In the following, we prove their conjecture about the groundstate, which states that the groundstates always have a form where the $-1$ entries occur first, and after the last $-1$, there are just $+1$ entries. 
First, note that their coupling matrices can be written as:
\begin{equation}
    A = \mathbb{E }- C - R,
\end{equation}
where 
\begin{align}
    \mathbb{E }_{i,j}&= 1 \\
    C_{i,j} &= \frac{i-1}{N-1} \\
        R_{i,j} &= \frac{j-1}{N-1}. 
\end{align}
Now consider the following 
\begin{align}
    &\argmin_{s \in \{ -1,1 \}}  s^T (\mathbb{E }- C - R)s + b^T s \\&= \argmin_{s \in \{ -1,1 \}}  s^T \mathbb{E }s- s^TCs - s^TRs + b^T s  \\
    &= \argmin_{s \in \{ -1,1 \}}  s^T \mathbb{E }s- s^TC^Ts - s^TRs + b^T s 
    \\
    &= \argmin_{s \in \{ -1,1 \}}  s^T (\mathbb{E }-2R)s+ b^T s.
\end{align}

Written out, one has to optimize
\begin{align}
&\sum_i (1- 2 \frac{i-1}{N-1})s_i + \sum_{i,j\quad i \neq j } (1-\frac{i+j-2}{N-1}) s_is_j\\
&=\sum_i (1- 2 \frac{i-1}{N-1})s_i + \sum_{i,j \quad i \neq j} (1-\frac{2i-2}{N-1}) s_is_j\\
  &=  \sum_i  s_i(1+ \sum_ {j\neq i } s_j)(1- \frac{2(i-1)}{N-1}).
\end{align}
Now we add the term $\sum_i  s_i^2 (1- \frac{2(i-1)}{N-1})$
, which is constant in $s_i$, since $s_i^2=1$.
This yields 

\begin{equation}
  (1+ \sum_ {j } s_j)  \sum_i  s_i(1- \frac{2(i-1)}{N-1})
\end{equation}
which we want to optimize.
Note first that the first factor only depends on the number of variables that are $+1$ and not on which ones have the value.
The second factor is maximized or minimized for 
\begin{equation}
    s_{+-}:= (1,1,,...,1,-1,...,-1)
\end{equation}
or
\begin{equation}
    s_{-+} := (-1,-1,,...,-1,1,...,1)= -s_{+-}.
\end{equation}

Additionally we know that 
\begin{equation}
    s^T A s = (-s)^TA(-s).
\end{equation}
This tells us that the question of whether $s$ or $-s$ has a better energy only depends on the linear term.
We will show that $b^T s_{+-}\geq 0$. This would then yield that $b^T s_{+-}\leq 0$ which concludes the proof that the minimizer is always of the form $s_{-+}$.
Let $k$ denote the index where the signs change in the vector, then
\begin{align}
    b^T s_{+-}&= \sum_{i=1}^k (1-2\frac{i-1}{N-1}) - \sum_{i= k+1}^{N} (1- 2 \frac{i-1}{N-1})  \nonumber  \\ 
    &=  k-2\frac{1}{N-1}(-k+\sum_{i=1}^k i) - (N-k-2\frac{1}{N-1}(-(N-k)+\sum_{i=k+1}^N i))  \nonumber \\
    &=  k-2\frac{1}{N-1}(-k+\frac{k(k+1)}{2}) - (N-k-2\frac{1}{N-1}(-(N-k) \nonumber \\ &+( \frac{N(N+1)}{2} -\frac{k(k+1)}{2})))   \nonumber \\
       &=  2(k-2\frac{1}{N-1}(-k+\frac{k(k+1)}{2})) - (N-2\frac{1}{N-1}(-N+( \frac{N(N+1)}{2} ))) \nonumber  \\
          &=  2(k-2\frac{1}{N-1}(\frac{k(k-1)}{2})) - (N-2\frac{1}{N-1}( \frac{N(N-1)}{2} ))  \nonumber \\
             &=  2(k-\frac{k(k-1)}{N-1}) - (N-N)  \nonumber \\
             &=  2k(1-\frac{k-1}{N-1}) \geq  0 \quad \forall k \in \{ 1,...,N \}.
\end{align}
Therefore, we know that vectors $s_{+-}$ have always worse or equal energy than vectors $s_{-+}$, and the complete statement from \cite{mehta2022hardness} is proven.
The problem instances which are publicly accessible from Ref.~\cite{mehta2022hardness} only are upper triangular, which is also indicated in their publication. But the proof works in the same way as for the convention with symmetric matrices by just introducing a factor $\frac{1}{2}$ in equation \eqref{eq:CouplingMehta}.

\section{Ablation Study}\label{sub:ablation}
\begin{figure}[H]
    \centering
    \includegraphics[trim={2cm 0cm 2cm 0cm},width= 0.5\textwidth]{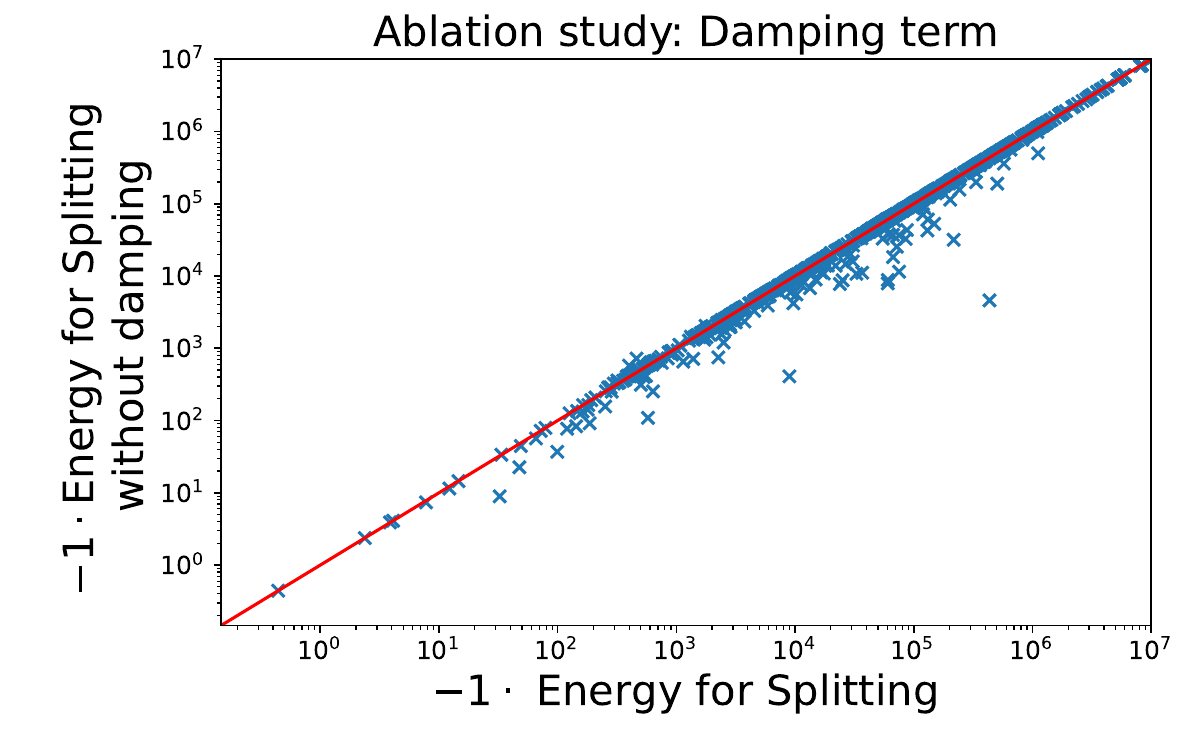}
    \caption{ Comparison of the final objective function values (energies) with the method without damping, i.e. $\lambda=0$. From the $922$ problem instances with sizes between $500$ and $1000$ vertices, in 156 cases both methods achieved the same result. In $383$ cases, the version with no damping was better and in $383$ cases, the version with damping was better. However, the method with damping is even better than the method without in  $296$ cases if you multiply it by a factor $1.001$. The reverse statement is only true in $164$ cases. This implies that for instances where no damping had a better result, the improvement is rather small. }
    \label{fig:AblationNoDamping}
\end{figure}

We will show an experiment to confirm that the damping term in equation \eqref{eq:Specific} improves the result. For this, we compare against the same method with the regularization parameter $\lambda$ set to zero. For the comparison method, where $\lambda$ is set to zero, we draw a new permutation in every subiteration. This is the case since in the original algorithm, we would only optimize the $\lambda $ parameter within the subiterations. The experiment on 922 Maximum cut problems can be seen in \cref{fig:AblationNoDamping}. For the $\lambda=0$ case, the lowest energy from all the subiterations was used.

\section{MQlib experiments with k-Opt} \label{sec:kOptExp}
\begin{figure}[H]
  \centering
 \includegraphics[trim={2cm 0cm 2cm 0cm},width=0.5\textwidth]{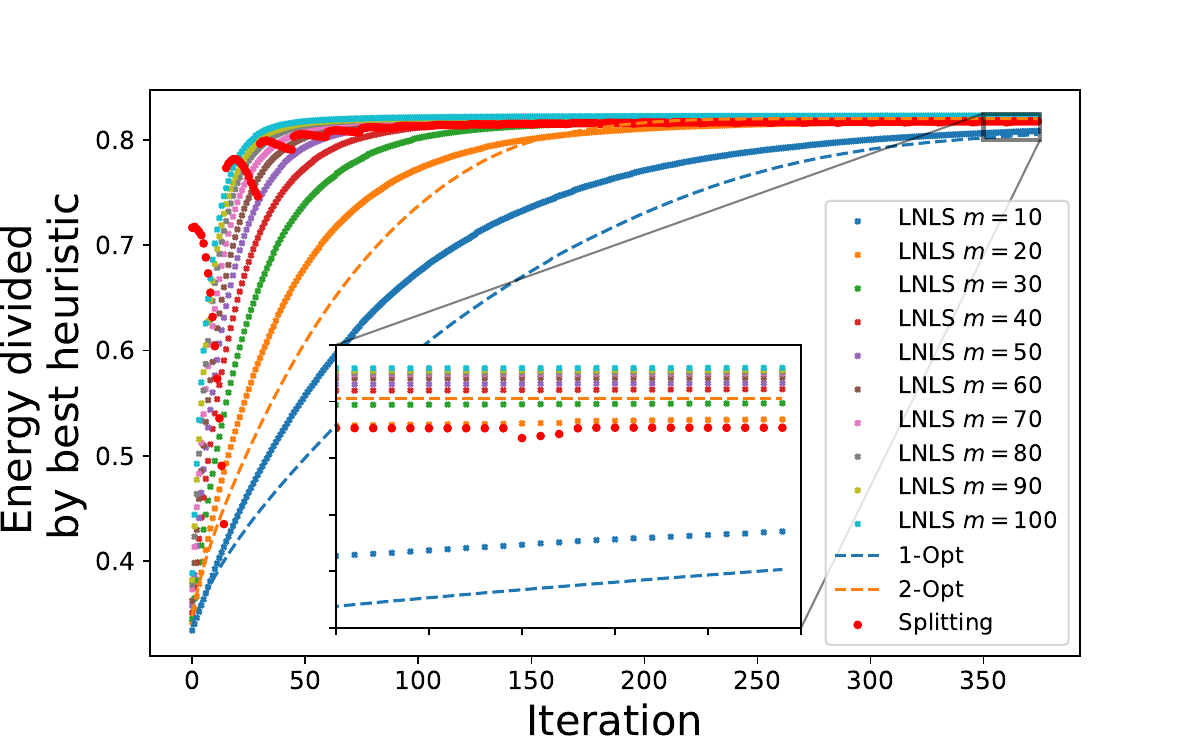}
  \caption{ Approximation ratio of 1-Opt and 2-Opt for $909$ MQLib instances included in \cref{fig: MainExpVerlauf}. }
  \label{fig:MQLibwithkOpt}
\end{figure}

For the $909$ MQlib instances that we investigated in \cref{sec:ExperimentsMQlib}, we also computed for 1-Opt and 2-Opt the average approximation ratios depending on the iteration number. This can be seen in \cref{fig:MQLibwithkOpt}.

\end{appendices}
\section*{Acknowledgment}
We thank Natacha Kuete Meli for careful proofreading of the manuscript and helpful discussion. 
MK is funded by Fujitsu Germany GmbH as part of the endowed professorship ``Quantum Inspired and Quantum Optimization''. 
\bibliography{sn-bibliography}

\end{document}